\documentclass[aps,pra,superscriptaddress,twocolumn,amsmath,amssymb,floatfix]{revtex4-2}

\pdfoutput=1
\usepackage{xcolor}
\definecolor{myblue}{rgb}{0.2,0.2,0.8}
\definecolor{myred}{rgb}{1,0.,0.3}
\usepackage{dcolumn}
\usepackage{bm}
\usepackage[colorlinks=true, citecolor=myblue, linkcolor=myred, urlcolor=myblue]{hyperref}
\usepackage{graphicx}
\usepackage{physics}
\usepackage{float}
\usepackage{tikz}
\usepackage{mathtools}
\usepackage{amsmath}
\usepackage{multirow}
\usepackage{dsfont}
\usepackage{tikz}
\usepackage{siunitx}
\usepackage{svg}
\usepackage{placeins}
\usepackage{soul}
\setstcolor{blue}
\usepackage{colortbl}
\usepackage{hhline}
\usetikzlibrary{positioning,calc,shapes,arrows,shapes.multipart,quantikz}

\newcommand{\rwth}{Institute for Quantum Information, RWTH Aachen University, 52056 Aachen, Germany}
\newcommand{\fzj}{Peter Grünberg Institute, Theoretical Nanoelectronics, Forschungszentrum Jülich, 52425 Jülich, Germany}
\newcommand{\lmu}{Fakultät für Physik, Ludwig-Maximilians-Universität München, 80799 München, Germany}
\newcommand{\mpq}{Max-Planck-Institut für Quantenoptik, 85748 Garching, Germany}
\newcommand{\mcqst}{Munich Center for Quantum Science and Technology (MCQST), 80799 München, Germany}

\begin{document}

\title{Bosonic Quantum Error Correction with Neutral Atoms in Optical Dipole Traps}

\author{Leon H. Bohnmann} \thanks{These authors contributed equally} \email[\\]{leon.bohnmann@rwth-aachen.de} \email{d.locher@fz-juelich.de} \affiliation{\rwth} \affiliation{\fzj}
\author{David F. Locher} \thanks{These authors contributed equally} \email[\\]{leon.bohnmann@rwth-aachen.de} \email{d.locher@fz-juelich.de} \affiliation{\rwth} \affiliation{\fzj}
\author{Johannes Zeiher} \affiliation{\lmu} \affiliation{\mpq} \affiliation{\mcqst}
\author{Markus Müller} \affiliation{\rwth} \affiliation{\fzj}


\date{February 27, 2025}

\begin{abstract}
Bosonic quantum error correction codes encode logical qubits in the Hilbert space of one or multiple harmonic oscillators. A prominent class of bosonic codes is that of Gottesman--Kitaev--Preskill (GKP) codes of which implementations have been demonstrated with trapped ions and microwave cavities.
In this paper, we investigate theoretically the preparation and error correction of a GKP qubit in a vibrational mode of a neutral atom stored in an optical dipole trap.
This platform has recently shown remarkable progress in simultaneously controlling the motional and electronic degrees of freedom of trapped atoms. 
The protocols we develop make use of motional states and, additionally, internal electronic states of the trapped atom to serve as an ancilla qubit.
We compare optical tweezer arrays and optical lattices and find that the latter provide more flexible control over the confinement in the out-of-plane direction, which can be utilized to optimize the conditions for the implementation of GKP codes. Concretely, the different frequency scales that the harmonic oscillators in the axial and radial lattice directions exhibit and a small oscillator anharmonicity prove to be beneficial for robust encodings of GKP states.
Finally, we underpin the experimental feasibility of the proposed protocols by numerically simulating the preparation of GKP qubits in an optical lattice with realistic parameters.
\end{abstract}

\maketitle

\section{Introduction}\label{sec:introduction}

\begin{figure*}[t]
    \includegraphics[width=0.99\linewidth]{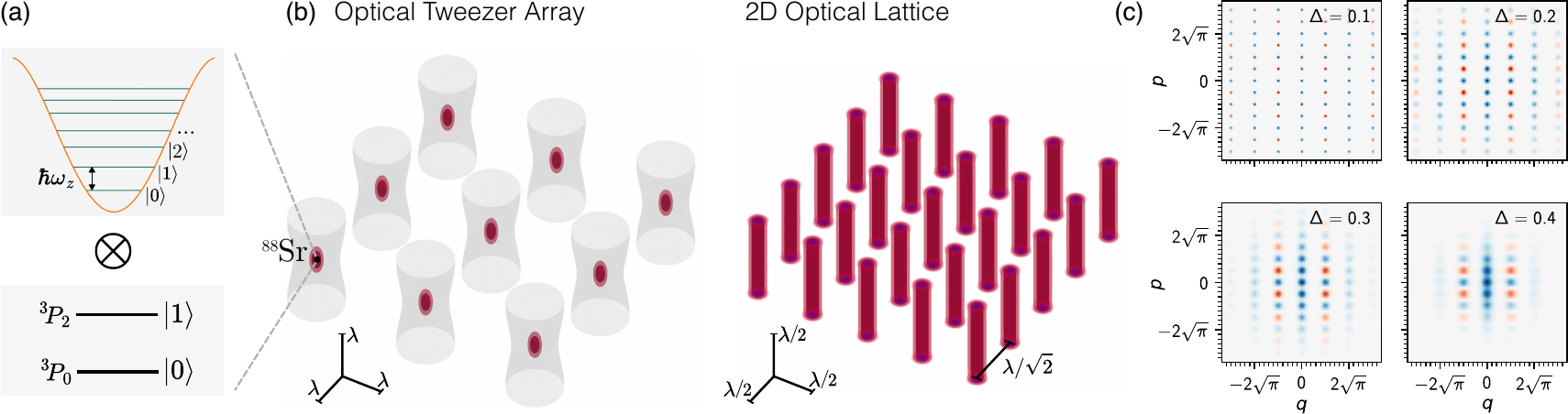}
    \caption{\textbf{Neutral atoms in optical dipole traps can encode GKP states.} (a)~An atom confined in an optical dipole trap exhibits vibrational modes. One of these bosonic modes shall be used to encode a GKP qubit. This requires that the bosonic degrees of freedom are coupled to an ancilla qubit formed by two electronic states of the trapped atom, e.g. the fine-structure qubit in~$^{88}\mathrm{Sr}$. (b)~Optical tweezer arrays and 2D optical lattices are both highly scalable setups. Atoms trapped in optical tweezers are confined to similar length scales in all three spatial dimensions. In optical lattices, the traps can have a tube-like shape with strong confinement in the radial (in-plane) directions and weak confinement in the axial (here vertical) direction. This is beneficial for the encoding of GKP states, since it allows one to realize squeezing in the radial mode only, while heating in the axial modes is suppressed. (c)~Visualization of finite square GKP code states. The Wigner function of the logical state $\ket{0_L^\Delta}$ is shown for different values of $\Delta$ according to Eq.~\eqref{eq:GKP finite is boson-loss}. The limit $\Delta \rightarrow 0$ corresponds to the ideal GKP code state as defined in Eq.~\eqref{eq:GKP ideal 0}.}
    \label{fig:title figure}
\end{figure*}

Fault-tolerant quantum computing requires the encoding of quantum information in logical qubits such that quantum error correction (QEC) can be performed repeatedly~\cite{terhal_2015_quantum, campbell_2017_road}.
Bosonic QEC codes, also known as continuous-variable codes, encode logical qubits in the Hilbert space of one or multiple harmonic oscillators~\cite{Albert:2018, Terhal:2020, cai:2021, albert:2022}. Multiple classes of bosonic codes have been proposed, such as cat codes~\cite{Cochrane:1999, Leghtas:2013b, Mirrahimi:2014, Ofek:2016}, binomial codes~\cite{Michael:2016, Hu_2019_quantum}, and so-called Gottesman--Kitaev--Preskill (GKP) codes~\cite{GKP:2001, Grimsmo:2021, brady_2024_advances}.
Logical states of GKP codes, on which we focus in the present paper, have been prepared experimentally in microwave cavities coupled to a transmon qubit~\cite{campagne:2020, Eickbusch:2022, Sivak:2023, Quirion:2024} and with trapped ions~\cite{Flühmann:2019, Neeve:2020, Matsos:2023}, and also QEC with these codes has been demonstrated in both platforms~\cite{campagne:2020, Sivak:2023, Neeve:2020, Quirion:2024}.
Another promising experimental platform that is well suited for quantum computing is that of neutral atoms trapped in optical lattices or optical tweezers~\cite{saffman_2010_quantum, saffman_2016_quantum, morgado_2021_quantum}.
Qubit-based quantum computing in these setups has recently shown great progress by demonstrating e.g. high-fidelity entangling gates and elements of quantum error correction and fault tolerance~\cite{Graham2022multi,Ma2023high,Evered2023high,Bluvstein2024logical}.
On the other hand, an atom confined by an optical dipole trap in the three spatial dimensions exhibits three bosonic modes, and in past years, precise control of the motional states of such trapped atoms has been demonstrated in experiments~\cite{Moringa:1999, Bouchoule:1999, Foerster:2009, Belmechri:2013, Winkelmann:2022, Hartke2022, Brown:2023, Scholl:2023}.
As an example, simultaneous entanglement of both motional and electronic states of two $^{88}\mathrm{Sr}$ atoms trapped in optical tweezers has been realized lately~\cite{Scholl:2023}.
This raises the question of whether and how GKP codes can be implemented in this platform as well, which has been studied in recent Refs.~\cite{Kendell:2023, grochowski2024} by means of trap potential modulation. To explore opportunities of this physical platform for bosonic QEC~\cite{liu2024hybrid}, the implementation of more general bosonic quantum circuits shall be analyzed in this paper.

In this paper we investigate the suitability of atoms trapped in optical tweezers and optical lattices to encode and operate GKP qubits.
Specifically, the atoms are confined in state-dependent optical dipole traps in order to couple the atomic motion to an ancilla qubit formed by two internal electronic states, as indicated in Fig.~\ref{fig:title figure}(a).
We find that anisotropic two-dimensional (2D) optical lattices are well suited for the encoding of GKP qubits because the harmonic oscillator frequency in one spatial dimension can be chosen to differ significantly from the frequencies of the other two spatial oscillator modes, as illustrated in Fig.~\ref{fig:title figure}(b). This allows one
to manipulate the relevant coding mode independently from the unused spectator modes.
Furthermore, the system is scalable and exhibits only a small oscillator anharmonicity, which deteriorates encoded states steadily.
The paper is structured as follows.
In Sec.~\ref{sec:gkp_theory} we provide an introduction to GKP codes and present protocols for preparation and error correction of GKP code states in general.
In Sec.~\ref{sec:gkp_preparation} we explain how the bosonic operations required to operate GKP codes, such as squeezing and conditional displacements, can be realized with neutral atoms in optical dipole traps.
In Sec.~\ref{sec:simulation} we analyze the suitability of optical tweezers and optical lattices for operation of GKP codes in detail.
We numerically simulate the preparation of a GKP code state in an optical lattice to demonstrate the experimental feasibility of the proposed protocols.
Finally, in Sec.~\ref{sec:conclusion} we summarize our results.

\section{GKP Codes}\label{sec:gkp_theory}
In 2001, Gottesman, Kitaev, and Preskill proposed a class of bosonic quantum error correction codes now called GKP codes~\cite{GKP:2001}, which are intensely studied nowadays.
It has been shown theoretically that they are well suited to correct for boson loss~\cite{Albert:2018,Noh:2019},
which is a predominant source of decoherence in microwave cavities.
In neutral atom platforms, heating deteriorates quantum information encoded in the motional state of an atom, which can also be corrected with GKP codes~\cite{brady_2024_advances}.
Recently, break-even error correction compared to the Fock-encoding, which identifies $\ket{0_L} \hat{=} \ket{0}$ and $\ket{1_L} \hat{=} \ket{1}$, has been achieved experimentally using a GKP code~\cite{Sivak:2023}. In this section we review the basic properties of GKP codes and outline a preparation scheme that is adapted to GKP states with a finite boson number.

\subsection{Ideal GKP Code}
Ideal GKP code states are grid-like superposition states in phase space, composed of infinitely many quadrature eigenstates.
In the context of GKP codes, the quadrature operators are defined as $\hat{q} = (\hat{a}^\dagger + \hat{a})/\sqrt{2}$ and $\hat{p} = i (\hat{a}^\dagger - \hat{a})/\sqrt{2}$, such that $[\hat{q},\hat{p}]= i$~\footnote{In quantum optics the alternative definition $\hat{q} = (\hat{a}^\dagger + \hat{a})/2$ and $\hat{p} = i (\hat{a}^\dagger - \hat{a})/2$ is common, giving rise to the relation $[\hat{q},\hat{p}]= i/2$. For GKP codes $[\hat{q},\hat{p}]= i$ is typically preferred, hence the definition above.}.
The square GKP code will work as an instructive example throughout this paper. Protocols similar to the ones discussed in the following can be devised for other types of GKP codes. The logical zero state of the square GKP code reads~\cite{GKP:2001}
\begin{equation} \label{eq:GKP ideal 0}
    \ket{0_L} = \sum_{k=-\infty}^{\infty} \ket{q\!=\!2k\sqrt{\pi}} = \sum_{k=-\infty}^{\infty} \ket{p\!=\!k\sqrt{\pi}} .
\end{equation}
The state can be interpreted as a superposition of position eigenstates or, equivalently, as a superposition of momentum eigenstates, which is a result of its grid-like appearance in phase space, shown in Fig.~\ref{fig:title figure}(c).
Ideal GKP states are an idealization, since they have an infinite expected boson number and, thus, they cannot be realized experimentally. Nevertheless, they build the foundation for experimentally accessible finite GKP codes.

GKP codes are stabilizer codes, which means that all code states are +1 eigenstates of a set of commuting operators which form the so-called stabilizer group.
We will see that specific displacements $\hat{D}(\alpha) \coloneq \exp[(\alpha \hat{a}^\dagger - \alpha^* \hat{a})/\sqrt{2}] = \exp[-i\Re(\alpha)\hat{p} + i\Im(\alpha)\hat{q}]$ serve as stabilizers.
Note that in quantum optics one usually defines $\hat{D}(\alpha) \coloneq \exp[\alpha \hat{a}^\dagger - \alpha^* \hat{a}]$ which displaces $\hat{a} \mapsto \hat{a}+\alpha$, while the convention for $\hat{D}$ that is typically used for GKP states displaces $\hat{q} \mapsto \hat{q}+\Re(\alpha)$ and $\hat{p} \mapsto \hat{p}+\Im(\alpha)$ and hence $\hat{a} \mapsto \hat{a}+\alpha/\sqrt{2}$.
In GKP codes, we have two stabilizer generators $\hat{\mathcal{S}}_X$ and $\hat{\mathcal{S}}_Z$, which generate an infinite group of stabilizers with elements $\hat{\mathcal{S}}_X^{n_x}\hat{\mathcal{S}}_Z^{n_z}$, where $n_x, n_z \in \mathbb{Z}$.
Choosing $\hat{\mathcal{S}}_X$ and $\hat{\mathcal{S}}_Z$ as orthogonal displacements (e.g. one as a displacement in $\hat{q}$ and the other one as a displacement in $\hat{p}$), one obtains a rectangular GKP code. A special case of this is the square GKP code considered in this paper, which has stabilizer generators
\begin{equation}
    \label{Square GKP stabilizers}
    \hat{\mathcal{S}}_X = \hat{D} \left(2\sqrt{\pi}\right), \quad
    \hat{\mathcal{S}}_Z = \hat{D} \left(2i\sqrt{\pi}\right) .
\end{equation}
The logical operators are
\begin{equation}
    \hat{\mathcal{X}}_L = \hat{D} \left(\sqrt{\pi}\right), \quad 
    \hat{\mathcal{Z}}_L = \hat{D} \left(i\sqrt{\pi}\right) ,
\end{equation}
thus, they correspond to displacements by half a stabilizer distance.
As usual, the Pauli-$Y$ operator is obtained as $\hat{\mathcal{Y}}_L = i\hat{\mathcal{X}}_L\hat{\mathcal{Z}}_L$.
The stabilizers commute with each other and with the logical operators and the logical operators mutually anticommute.
This can be derived from the braiding relation~\cite{GKP:2001}
\begin{equation}
\label{eq:braiding relation}
    \hat{D}(\alpha)\hat{D}(\beta) = e^{-iA(\alpha, \beta)}\hat{D}(\beta)\hat{D}(\alpha) ,
\end{equation}
where $A(\alpha, \beta) = \Re(\alpha)\Im(\beta)-\Im(\alpha)\Re(\beta)$ is the area in phase space spanned by the two displacements.
For two commuting operators $A$ is an integer multiple of $2\pi$, whereas for anticommuting operators it is an odd integer multiple of $\pi$.

\subsection{Finite GKP Code}
For realistic physical systems finite approximate versions of GKP codes have been formulated~\cite{Matsuura:2020}.
One elegant form can be constructed by suppressing high boson number contributions exponentially~\cite{Terhal:2020}:
\begin{equation} \label{eq:GKP finite is boson-loss}
    \ket{\psi_L^{\Delta}} = 2\sqrt{\pi}\Delta e^{-\Delta^2\hat{a}^\dagger \hat{a}}\ket{\psi_L}.
\end{equation}
Some resulting finite GKP states are depicted in Fig.~\ref{fig:title figure}(c) for different values of $\Delta$. Most notably, for $\Delta \rightarrow 0$ the ideal GKP state is recreated and for $\Delta \rightarrow \infty$ the state reduces to the vacuum state.
Inserting the logical state $\ket{0_L}$ from Eq.~\eqref{eq:GKP ideal 0} into Eq.~\eqref{eq:GKP finite is boson-loss} and considering small values of $\Delta$ leads to the instructive expression
\begin{equation} \label{eq:GKP superposition of squeezed states}
    \ket{0_L^{\Delta}} \propto \sum_{k=-\infty}^{\infty} e^{-2\pi \Delta^2 k^2} \hat{D}\left(2k\sqrt{\pi}\right) \hat{S}\left(-\ln\Delta\right)\ket{0} + \mathcal{O}\left(\Delta^4\right) ,
\end{equation}
where $\hat{S}(z) = \exp[\frac{1}{2}(z^*\hat{a}^2 -  z\hat{a}^{\dagger 2})]$ denotes the squeeze operator.
We see that this finite GKP state is a superposition of squeezed states where the weights decay with displacement distance.
The derivation of this expression is shown in Appendix~\ref{appendix: finite GKP}.

For $\Delta \ll 1$ the parameter $\Delta$ can be identified with the \emph{effective squeezing}. Given a state $\hat{\rho}$ it is calculated as follows~\cite{Terhal:2020}:
\begin{equation}
    \Delta_{X,Z} \coloneq \sqrt{\frac{1}{2\pi} \ln\left(\frac{1}{|\text{Tr}(\hat{\mathcal{S}}_{X,Z}\hat{\rho})|^2}\right)}.
\end{equation}
The name \emph{effective squeezing} is motivated by squeezed vacuum states, since a state $\hat{S}(-\ln\Delta)\ket{0}$ has effective squeezings $\Delta_Z = \Delta$ and $\Delta_X = 1/\Delta$.
For general states, the values $\Delta_Z$ and $\Delta_X$ are not the multiplicative inverse of each other and for GKP states both are smaller than 1, which is why GKP states can be interpreted as squeezed in both $\hat{p}$ and $\hat{q}$.
For $\Delta_{X,Z} \rightarrow 0$ one obtains the ideal GKP code.
In experiments GKP states with effective squeezings in the range 0.30 -- 0.53 have been prepared with trapped ions~\cite{Flühmann:2019, Neeve:2020} and GKP states with effective squeezings of 0.34 have been demonstrated with microwave cavities coupled to a transmon qubit~\cite{Sivak:2023}.
Since $\Delta_X$ and $\Delta_Z$ behave multiplicatively during concatenation of squeeze operations, they are sometimes quantified in \si{dB} as $10\log_{10}\left(1/\Delta_{Z,X}^2\right)$.

\subsection{Preparation of GKP States} \label{sec:II.C GKP state prep}
\begin{figure} \centering
    \includegraphics[width=0.99\linewidth]{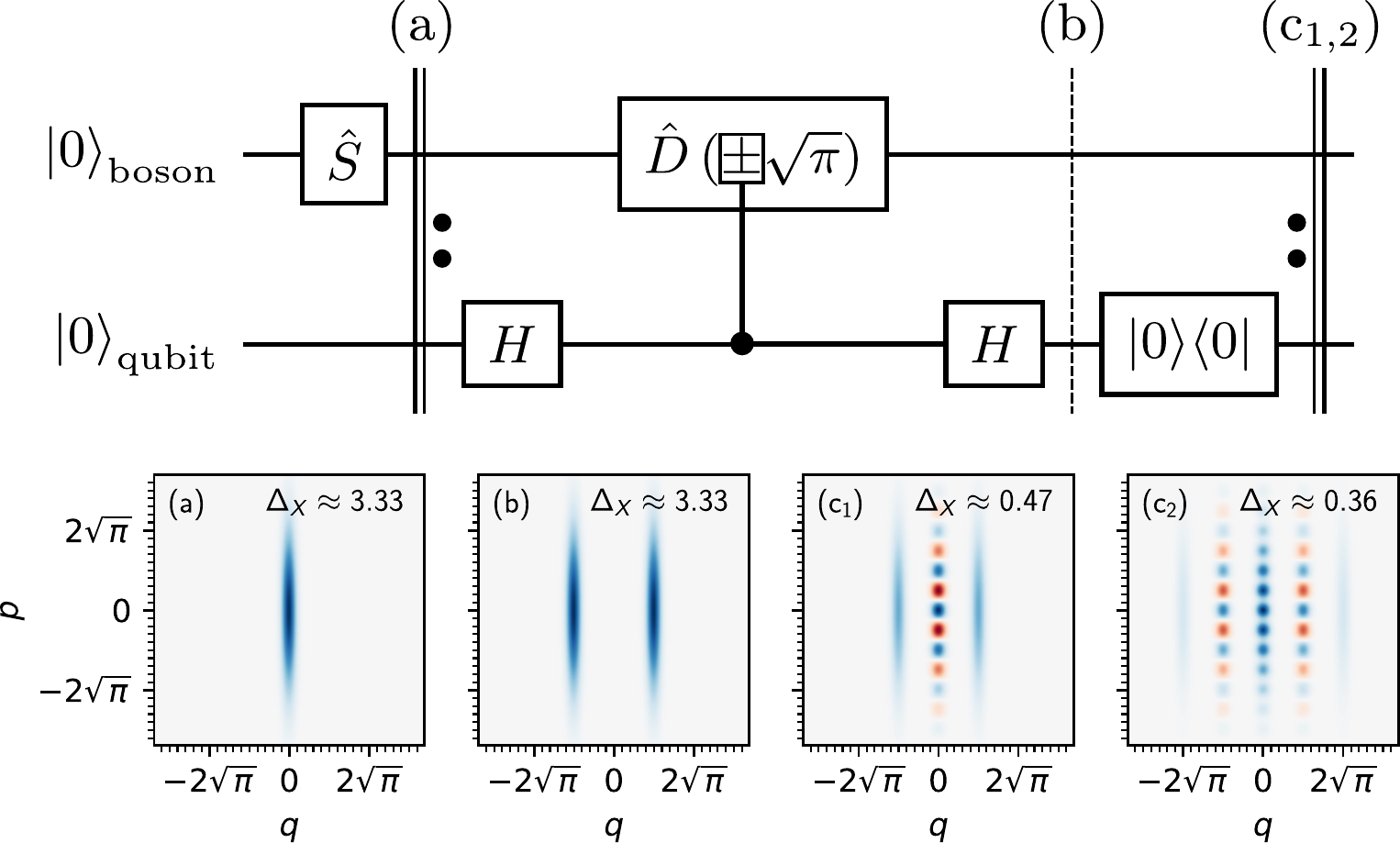}
    \caption{\textbf{Preparation of finite GKP states by repeated displacements and postselection.} Top: GKP preparation circuit in terms of the bosonic operations \emph{Squeeze} ($\hat{S}$) and \emph{Displace} ($\hat{D}$) and qubit operations \emph{Hadamard} ($H$) and projective postselection ($\ket{0} \!\bra{0}$). Bottom: Wigner functions of the bosonic subsystem at selected times during the first two preparation cycles. First, a squeezed state ($\Delta \coloneq \Delta_{Z}=0.3$, and $\Delta_{X} = 1/\Delta \approx 3.33$) is created, as shown in (a). This state is an approximate eigenstate of the stabilizer $\hat{\mathcal{S}}_Z$. The state is partially displaced to the right and partially to the left by an ancilla-conditioned application of half a stabilizer distance. This is equivalent to an operator measurement of $\hat{\mathcal{S}}_X$ combined with an unconditional application of \mbox{$\hat{\mathcal{S}}_X^{-1/2} \hat{=} \, \hat{\mathcal{X}}_L$} to recenter the state. 
    The bosonic state is still entangled with the ancilla, thus the bosonic state obtained when tracing over the ancilla appears as a mixture of two squeezed states, see (b). After the application of a Hadamard gate the ancilla is projected onto the state $\ket{0}$, thus effectively postselecting on $\ket{+}$, such that the bosonic state is projected onto a superposition of two squeezed states, depicted in (c\textsubscript{1}). A second round of conditional displacement and postselection results in a superposition of three squeezed states $[\hat{D}(-2\sqrt{\pi}) + 2\hat{D}(0) + \hat{D}(2\sqrt{\pi})] \hat{S} \ket{0}$, shown in (c\textsubscript{2}).}
    \label{fig:GKP preparation repetition wigner}
\end{figure}
In Eq.~\eqref{eq:GKP superposition of squeezed states} in the previous subsection we have seen that finite GKP codes can be understood as superpositions of squeezed states, where squeezed states that are closer to the origin have a higher contribution.
For the preparation of finite GKP states one can thus start with a squeezed vacuum state and create a superposition of displaced versions of it. To create these superpositions, all methods presented will use an additional two-level system, referred to as the ancilla qubit, which can be coupled to the bosonic degrees of freedom.
In a neutral atom two internal electronic states can be used to form the ancilla qubit, as indicated in Fig.~\ref{fig:title figure}(a) and discussed in more detail in Sec.~\ref{sec:gkp_preparation}.
Here we outline two protocols for GKP state preparation: The first one uses postselection on mid-circuit ancilla measurements while the second one makes use of a corrective displacement and an ancilla reset operation. The second scheme can also be used for error correction with adapted parameters.
\begin{figure*}
    \includegraphics[width=0.99\linewidth]{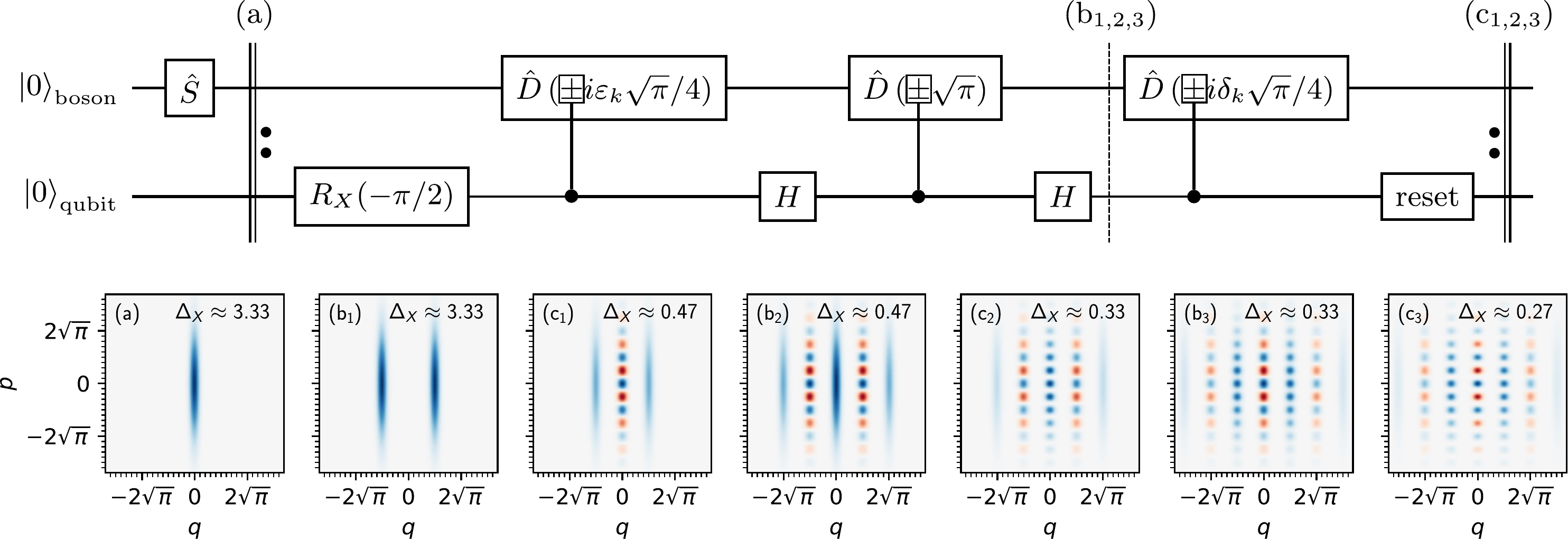}
    \caption{\textbf{Deterministic preparation of finite GKP states.} Top: GKP state preparation and error correction circuit, consisting of a squeeze operation ($\hat{S}$), conditional pre-rotation $\hat{D}\left(\pm i\varepsilon \sqrt{\pi}/4\right)$, stabilizer application $\hat{D}\left(\pm \sqrt{\pi}\right)$ and correction $\hat{D}\left(\pm i\delta \sqrt{\pi}/4\right)$. Each stabilizer application includes an unconditional application of $\hat{\mathcal{X}}_L$, such that the resulting state after three cycles is of type $\ket{1_L^\Delta}$. The required operations on the ancilla qubit are $X$-rotations $R_X(\theta) = \exp(-i \theta \sigma_x / 2)$, where $\sigma_x$ denotes the Pauli $X$ matrix, Hadamard gates $H$ and qubit reset. Bottom: Wigner function of the bosonic subsystem at selected points in time during the first three state preparation cycles with $\varepsilon_k=0$ and $\delta_1=1$, $\delta_2=0.5$, $\delta_3\approx0.31$.}
    \label{fig:GKP sharpen circuit}
\end{figure*}

\subsubsection{Postselection Scheme}

The postselection scheme is shown in Fig.~\ref{fig:GKP preparation repetition wigner}.
A variation of it was formalized in Ref.~\cite{Travaglione:2002} and it has been realized experimentally with trapped ions in Ref.~\cite{Flühmann:2019}. We assume that a harmonic oscillator and a qubit have been initialized in their ground states. On the harmonic oscillator, a single squeeze operation is needed, which also determines the final effective squeezing $\Delta_Z$ up to experimental imperfections. This operation also \emph{increases} $\Delta_X$ to $\Delta_X = 1/\Delta_Z$, which then has to be lowered during the main procedure. The main procedure repeatedly splits the squeezed states into two displaced copies. The Hadamard gate prepares the ancilla qubit in the state $\ket{+} = (\ket{0} + \ket{1})/\sqrt{2}$. Then, a conditional displacement $\hat{D}(\sqrt{\pi}) \otimes \ket{0}\!\bra{0} + \hat{D}(-\sqrt{\pi}) \otimes \ket{1}\!\bra{1}$ is applied
and, finally, the protocol postselects on the positive superposition of both paths. The two Hadamard gates on the ancilla qubit ensure the basis change from $\ket{0}$/$\ket{1}$ to $\ket{+}$/$\ket{-}$ for the preparation and postselection. This procedure can also be interpreted as a repeated stabilizer measurement and projection onto the code space. The only difference is an included unconditional application of $\hat{\mathcal{X}}_L \hat{\mathcal{S}}_X^{-1}$, which changes the Kraus operator from $(\mathds{1}+\hat{\mathcal{S}}_X)/2$ to $\hat{K}_0 = [\hat{D}(-\sqrt{\pi}) + \hat{D}(\sqrt{\pi})]/2$. This difference ensures that the state stays centered around the origin after each cycle.
From the second round on, all new displaced states except for the leftmost and rightmost one interfere with one other state, for example after two rounds one obtains the Kraus operator $\hat{K}_{00} = \hat{K}_0^2 = [\hat{D}(-2\sqrt{\pi}) + 2\hat{D}(0) + \hat{D}(2\sqrt{\pi})]/4$ that contains three displacements. The weighting of the states follows a binomial distribution due to the binary random walk~\cite{Venegas:2012}, which converges to a Gaussian envelope as in Eq.~\eqref{eq:GKP superposition of squeezed states}.
The state preparation scheme introduced in Ref.~\cite{Travaglione:2002} measures higher orders of stabilizers which results in equal-weight superpositions of displaced squeezed states. These are examples of a more general technique called
\emph{linear combination of unitaries} (LCU)~\cite{Childs:2012, Omanakuttan:2023, Endo:2024}.

\subsubsection{Corrective Displacement Scheme}
The previous scheme heavily relies on postselection, which prohibits large scale parallel state preparation.
Moreover, measuring the electronic state of a trapped neutral atom would destroy the prepared motional state because of the large number of scattered optical photons and associated recoil heating during fluorescence imaging.
An approach to circumvent measurements and postselection is a correction $\hat{C}=\hat{D}(i\sqrt{\pi}/2)$ that is applied conditionally on the ancilla qubit being in state $\ket{1}$, instead of discarding that outcome. This correction anticommutes with the stabilizer through the braiding relation (Eq.~\eqref{eq:braiding relation}), such that it maps the previously discarded Kraus operator $\hat{K}_1\!=\!(\mathds{1}-\hat{\mathcal{S}})/2$ to the required one, $\hat{K}_0\!=\!(\mathds{1}+\hat{\mathcal{S}})/2$:
\begin{equation} \label{eq:GKP correction works once}
    \hat{C}\hat{K}_1 \ket{q\!=\!0} = \hat{K}_0\hat{C}\ket{q\!=\!0} = \hat{K}_0 \ket{q\!=\!0}.
\end{equation}
The last step relies on the fact that the state before the round is an approximate eigenstate of the corrective displacement. This is only true for position eigenstates and approximately true for a squeezed state, but it is no longer the case after the first cycle. To still use this approach, the stabilizer application can be replaced by a channel with Kraus operators $\hat{K}_\circlearrowleft\!=\!(\mathds{1}+i\hat{\mathcal{S}})/2$ and $\hat{K}_{\circlearrowright}\!=\!(\mathds{1}-i\hat{\mathcal{S}})/2$. Here, both outcomes can be corrected with a smaller corrective displacement $\hat{C}^{\pm\delta_k/2}=\hat{D}(\pm i\delta_k\sqrt{\pi}/4)$.
This approach is adapted from a scheme devised for error correction of finite GKP states~\cite{Royer:2020}, where it realizes a compromise of just a weak perturbation of the pre-procedure state and inclusion of information from the error correction round.
We numerically optimize the parameters $\delta_k$ to achieve a minimal $\Delta_X$ and find optimal values $\delta_1=1$, $\delta_2=0.5$, $\delta_3\approx0.31$, $\delta_4\approx0.217$, and $\delta_5\approx0.167$ when starting from a squeezed state with $\Delta_Z = 0.3$.
Note that the obtained values of $\delta_k$ are larger than the ones one would use for the error correction procedure because we cope with large initial values of $\Delta_X$.
This results in a complete preparation procedure to reach small values of $\Delta_X$.
Note that this protocol always produces statistical mixtures of states with different values of $\Delta_X$, since the initial squeezing does not prepare a perfect quadrature eigenstate.
Figure~\ref{fig:GKP sharpen circuit} shows the quantum circuit which, for $\varepsilon=0$, can be used for the preparation of GKP code states.
In Appendix~\ref{appendix: prepared states explicitly} we explicitly derive the approximate GKP states prepared in the first three cycles.
A similar approach has been proposed in Ref.~\cite{Hastrup:2021} and used in Ref.~\cite{Neeve:2020} for state preparation, using integer multiples of the stabilizer displacement distance instead of only $\hat{D}(\pm\sqrt{\pi})$ and finding analytical expressions for $\delta_k$ in that setup.
Note that the qubit reset of a neutral atom requires scattering of just a few photons. This procedure is therefore less detrimental for the encoded motional state than a qubit measurement.

The procedure outlined above can be adapted to work for error correction instead of state preparation.
Since in any experimental setup encoded GKP states continuously undergo incoherent noise processes, such as boson loss, heating, or oscillator dephasing, one has to perform QEC cycles repeatedly in order to preserve the encoded quantum information.
In the scheme above, in each round the state is partially displaced to a higher position quadrature, thus it should not be repeated infinitely often on an oscillator with an upper limit on the occupation number. To prevent this, de Neeve \textit{et al.}~\cite{Neeve:2020} introduced an ancilla pre-rotation step (see the first displacement in Fig.~\ref{fig:GKP sharpen circuit}). It has the effect that the stabilizer application does not displace the state symmetrically inwards and outwards, but with a bias towards the center, as detailed in Ref.~\cite{Neeve:2020}. This bias is achieved by rotating the ancilla conditioned on the position quadrature by an angle $\hat{q}\sqrt{\pi}\epsilon/4$. This is equivalent to a conditional displacement $\hat{D}(\pm i\varepsilon_k\sqrt{\pi}/4)$ like in the correction step, so it can be realized with the same techniques.

As a side remark, the complete correction scheme can alternatively be derived formally from a dissipative map which causes a decay towards the ground state of an unphysical Hamiltonian proportional to $\ln[\exp(-\Delta^2 \hat{n})\hat{\mathcal{S}}_X \exp(\Delta^2 \hat{n})]$, as shown in Ref.~\cite{Royer:2020}.
The operator $\exp(-\Delta^2 \hat{n})\hat{\mathcal{S}}_X \exp(\Delta^2 \hat{n})$ is an exact stabilizer for a finite GKP state $\ket{\psi^\Delta_L}$ with fixed $\Delta$. 
Through Trotterization of such a dissipator, ideal displacement distances $\epsilon=\delta= \sinh(\Delta^2)$ and a stabilizer application distance increased by a factor of $\cosh(\Delta^2)$ can be derived.

\section{Preparation of GKP States in Neutral Atoms}\label{sec:gkp_preparation}
Neutral atoms can be confined in optical dipole traps, whose most prominent representatives are optical tweezers and optical lattices~\cite{phillips_1998_nobel, grimm_2000_optical, Bloch:2005, kaufman_2021_quantum}.
Close to the trap minima, the motion of single atoms can to a good approximation be described by a three-dimensional harmonic oscillator. Any of the three directions can be used to encode a GKP state, even though some of them show favorable properties over the others. In the previous section we saw that an additional qubit degree of freedom that can be coupled to the motional state is required. For this we will dedicate two electronic levels of the atom.
We begin this section by recalling how a harmonic oscillator Hamiltonian arises for a neutral atom trapped by a Gaussian laser beam. Then we explain how squeeze operations and conditional displacements required for the preparation of GKP states can be realized in this platform.
Details on the ancilla qubit operations, such as single-qubit gates or reset, are not yet considered; we just ensure that in the composed protocols enough time is reserved to implement them.

\subsection{Optical Dipole Traps}
\label{sec:optical dipole traps}

A laser which is far-detuned from an atomic transition can cause a decrease in energy of one or several atomic levels due to the AC Stark effect~\cite{Foot:2005}. This decrease in energy is proportional to the light intensity~$I$. Under the assumption that the electronic dynamics are much faster than the atomic motion, the intensity-dependent energy thus acts as a potential $U(x) \propto -I(x)$ for the atom.
In Appendix~\ref{appendix: optical dipole traps} we provide a short introduction to optical dipole traps.
The potential $U(x)$ in general depends on the atom's electronic state.
A crucial requirement, however, is that both qubit states of the atom experience the same trapping potential, to avoid unintentional coupling of the ancilla qubit with the bosonic degrees of freedom. Therefore, one chooses the trap laser to work at a so-called magic wavelength~\cite{katori_1999_optimal, Katori_2011_optical} for the specific ancilla qubit. At exactly this wavelength, the trap potential is equal for the atom in either of the electronic ancilla states.

In a typical dipole trap, for example, a radial mode of an optical tweezer, the intensity profile and thus the potential can be modeled by a Gaussian
\begin{equation} \label{eq:intensity taylor}
    U(x) = -U_0 e^{-2\frac{x^2}{w_0^2}} = -U_0 + 2 \frac{x^2}{w_0^2} U_0 + \mathcal{O}(x^4) ,
\end{equation}
where $U_0$ is the trap depth and $w_0$ describes the beam width.
Close to the trap bottom, the potential can be expanded up to second order, yielding a harmonic oscillator potential.
In Sec.~\ref{sec:simulation} we analyze the potentials of realistic dipole traps, including higher-order anharmonic terms, while in this section the harmonic approximation is used to make the preparation scheme explicit.
Combining the harmonic potential with the kinetic energy of the atom gives rise to the Hamiltonian of a quantum harmonic oscillator:
\begin{equation} \label{eq:hamonic oscillator pure}
    \hat{H} = \frac{\hat{P}_x^2}{2m} + \frac{1}{2} m \omega^2 \hat{x}^2 ,
\end{equation}
where $\hat{x} = \sqrt{\hbar / (m \omega)} \, \hat{q}$ and $\hat{P}_x = \sqrt{\hbar m \omega} \, \hat{p}$ are the physical position and momentum operators of the atomic center of mass. For the Gaussian beam in Eq.~\eqref{eq:intensity taylor} the oscillator frequency is calculated to be
\begin{equation} \label{eq:frequency_potential_relation}
    \omega = \frac{2}{w_0} \sqrt{\frac{U_0}{m}}.
\end{equation}
For beam profiles that deviate from a Gaussian, $w_0$ has to be replaced by a similar characteristic length scale, as will be discussed in Sec.~\ref{sec:tweezer vs lattice}.

\subsection{Squeezing} \label{sec:squeezing}

\begin{figure}
    \centering
    \includegraphics[width=0.99\linewidth]{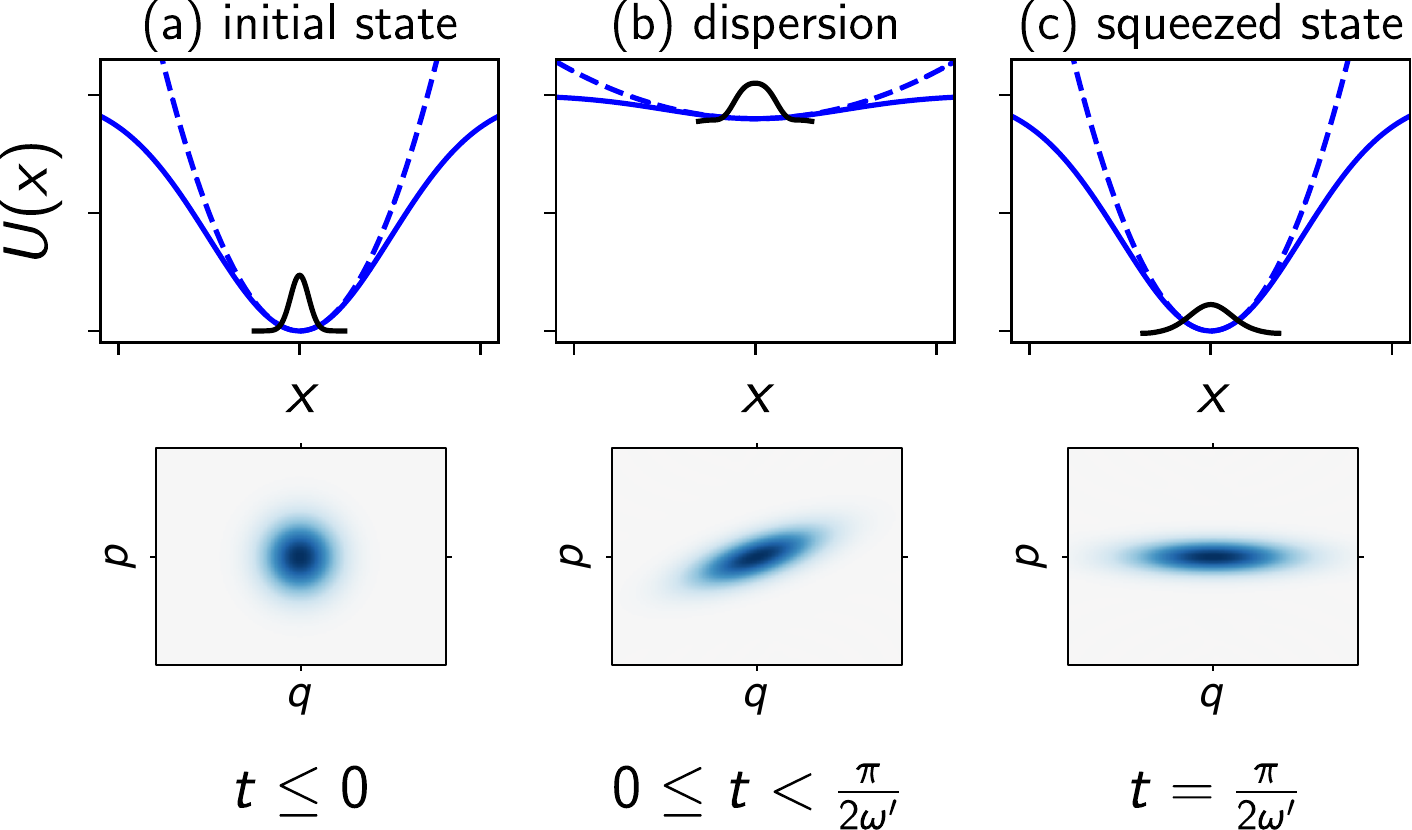}
    \caption{\textbf{Squeezing the motional state of a neutral atom.} Panel (a) shows a Gaussian trap potential of depth $U_0$ (solid blue line), its harmonic approximation (dashed blue line), as well as a sketch of a wave packet (black) and the Wigner function corresponding to the vacuum state of the harmonic oscillator. Changing the laser intensity and thus the potential depth rapidly from $U_0$ to $U'_0$ results in a change of the oscillator frequency from $\omega$ to $\omega'$. This causes the wave packet to disperse, as shown in panel (b). After a quarter of an oscillator period, $t = \pi / (2 \omega')$, the original potential depth is restored, as indicated in panel (c). The resulting state is a momentum-squeezed state.}
    \label{fig:tweezer squeezing demonstration}
\end{figure}

A squeeze operator is naturally generated by a Hamiltonian proportional to $\hat{x}^2$. Thus, an unconditional squeezing can be achieved by quickly changing the potential depth $U_0$ to a value $U'_0$~\cite{Brown:2023}. The resulting change in potential curvature changes the oscillator frequency $\omega \propto \sqrt{U_0}$ according to Eq.~\eqref{eq:frequency_potential_relation}.
The oscillator motion switches from a circular motion to a slower (or faster) \emph{elliptic} rotation at a frequency $\omega'$, as pictured in Fig.~\ref{fig:tweezer squeezing demonstration}.
After a rotation of $\pi/2$ in the elliptic oscillator, one obtains a squeezed state 
\begin{equation}
    \hat{S}\big(\ln\left(\omega'/\omega\right)\big)\ket{0}
\end{equation}
when starting in the bosonic vacuum $\ket{0}$. Note that this procedure is specifically engineered to transform the vacuum into a squeezed state. A derivation and general description as a unitary are sketched in Appendix~\ref{appendix: squeezing, displacement}. In general, states also accumulate a dynamical phase compared to an idling oscillator due to the modified rotation speed.
Note that this scheme has so far considered an isolated bosonic mode although one has to keep in mind that with a change of beam intensity the oscillator frequencies of all three spatial vibrational modes are changed. In Appendix~\ref{appendix: squeezing, displacement} we study how one can squeeze just the $z$-mode of an anisotropic three-dimensional harmonic oscillator with $\omega_z \ll \omega_{x,y}$. This can be achieved by adjusting the potential depth at a rate that is fast compared to $\omega_z$ but appears adiabatic to the $x$- und $y$-mode.

\subsection{Displacement}
\begin{figure}
\centering
\includegraphics[width=0.98\linewidth]{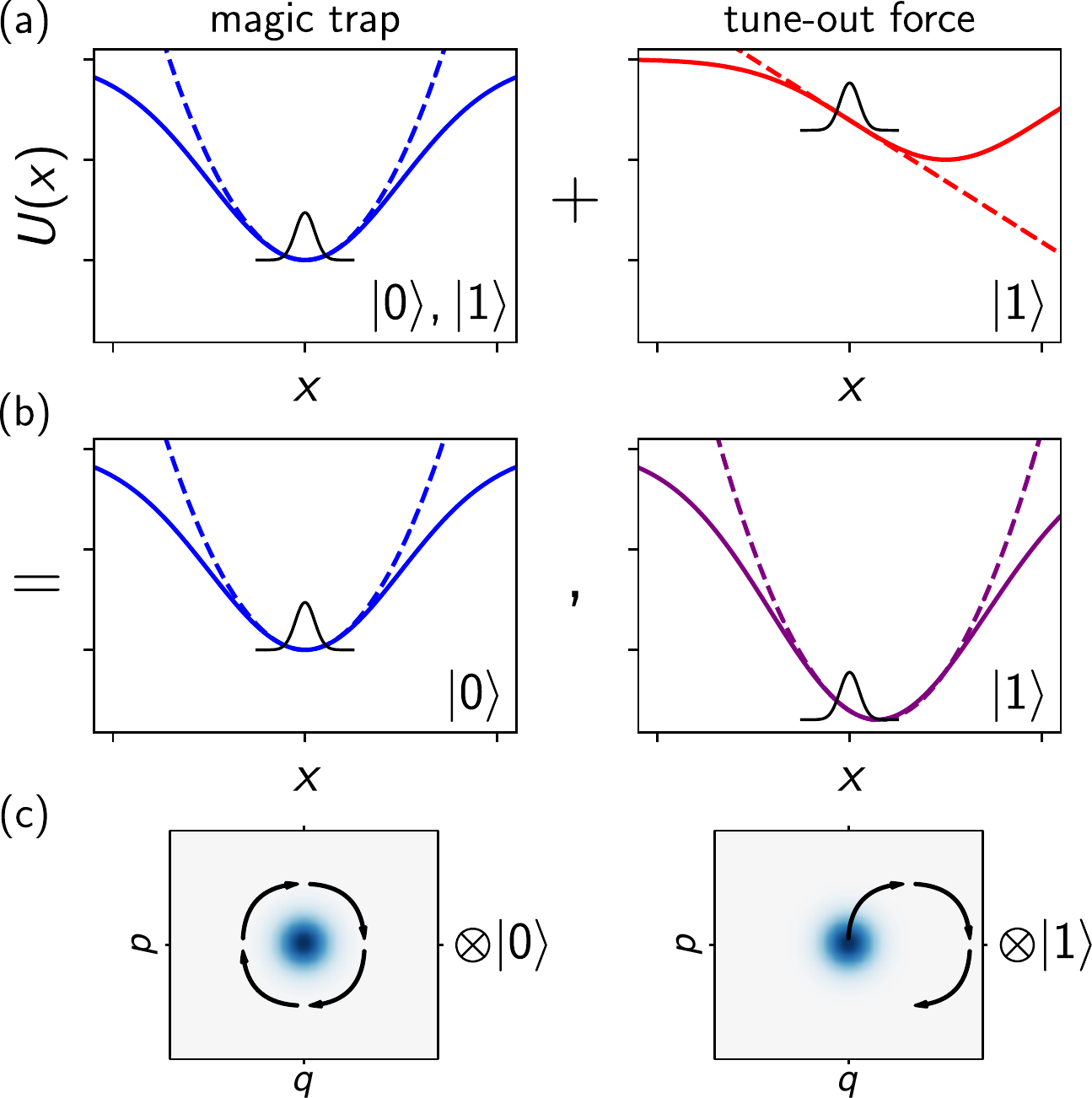}
    \caption{\textbf{Conditional displacement of the motional state of a neutral atom.} Panel (a) shows one-dimensional cuts of the trap potentials for the magic wavelength trap (blue) and the tune-out wavelength trap (red) realizing a state-dependent potential. Solid lines correspond to the Gaussian potentials, while dashed lines indicate harmonic approximations in the center. Note that the tune-out potential is placed such that the quadratic term vanishes. Panel (b) shows the combined potentials which an atom experiences in the two electronic qubit states $\ket{0}$ and $\ket{1}$. Both oscillators have approximately the same frequency $\omega$ but different center points. Panel (c) shows the dynamics of a Fock state under this potential, which is a conditional displacement combined with an unconditional rotation in phase space, see Eq.~\eqref{eq:force unitary}. Combining this operation with an idling rotation realizes a conditional displacement according to Eq.~\eqref{eq:Momentum kick}.}
    \label{fig:tweezer addition demonstration}
\end{figure}

For the protocols described in Sec.~\ref{sec:gkp_theory}, ancilla-conditioned displacements are required. An unconditional displacement $\hat{D}(\alpha)$ can be realized by moving the trap center, and thus the reference frame rapidly by $d=-\alpha \Delta x$, where $\Delta x = \sqrt{\hbar/(m\omega)}$ is the harmonic oscillator length.
Alternatively, one can use an additional laser beam operating at the magic wavelength to realize a force $-f \hat{x}$ and thus shift the trap center.
A conditional displacement must end in a shared reference frame for the displaced and non-displaced path. Thus, the effective trap center can be moved through a state-dependent force term $-f \hat{x} \otimes \ket{1}\!\bra{1}$ added to the Hamiltonian and then be moved back by disabling the force~\cite{Leibfried:2003}.
For simplicity we consider constant forces that are switched on and off instantaneously.
Such a state-dependent force can be realized with an additional laser operating at a so-called tune-out wavelength~\cite{leblanc_2007_species}. At this wavelength one of the qubit levels experiences no AC Stark shift while the other one experiences a non-zero shift.
The application of a constant force $f$ corresponds to a displacement of the trap center by a value
\begin{equation}
    \alpha_{d}  = \frac{f}{\sqrt{\hbar m \omega^3}}
\end{equation}
in phase space.
In a duration $t$ this realizes the unitary 
\begin{equation} \label{eq:force unitary}
    \hat{U}(t) = \hat{D}\!\left(\alpha_d\left[1-e^{-i\omega t}\right]\right) \hat{R}\left( \omega t \right) e^{i\theta(t)},
\end{equation}
for an atom in the electronic state $\ket{1}$, depicted in Fig.~\ref{fig:tweezer addition demonstration}.
This is not a pure displacement but includes a rotation $\hat{R}(\omega t) = e^{-i\omega t\hat{a}^\dagger \hat{a}}$ and the acquisition of a phase $\theta(t) = \frac{1}{2} \alpha_{d}^2 \left[ \omega t - \sin(\omega t) \right]$ which only matters if the operation is conditioned on the ancilla qubit.
If one includes equal waiting times before and after the displacement pulse, such that the procedure takes a full oscillator cycle, the \textit{direction} of the displacement no longer depends on the displacement duration $t$:
\begin{equation} \label{eq:Momentum kick}
\begin{aligned} 
     \hat{R}\left(\pi-\omega t/2\right) \; &
     \hat{U}(t) \,
     \hat{R}\left(\pi-\omega t/2\right) \\
     & = \hat{D}\big(-2i\alpha_d \sin(\omega t/2)\big) e^{i\theta(t)}.
\end{aligned}
\end{equation}
The distance of the displacement can then be controlled by the two parameters $f$ and $t$, while the direction can be controlled independently by choosing an appropriate starting time of the whole procedure. In Appendix~\ref{appendix: squeezing, displacement} we derive these relations in more detail.

The force can be realized by placing a Gaussian laser beam of width $w_1$ and depth $U_1$ which operates at the tune-out wavelength at a distance of $w_1/2$ next to the trap center. At exactly this distance the curvature is zero such that the harmonic part of the potential is neither increased nor decreased.
The atom then experiences a force $f=2 e^{-1/2} U_1 /w_1$ and its potential energy is lowered by an amount $e^{-1/2} U_1$.
This lowering in energy does not affect the bosonic mode but acts as a phase feedback $e^{-1/2} U_1 t / \hbar$ on the ancilla qubit which is added to $\theta(t)$. The phase must either be fine-tuned to a multiple of $2\pi$ or corrected through an ancilla-only operation. Methods from optimal control or pulse echoing may also prove useful for this fine-tuning~\cite{Werschnik_2007}.
The beam waist $w_1$ should be as large as possible because the unwanted higher orders scale with $1/w_1^3$ while the gradient and thus displacement speed only decreases with $1/w_1$. These issues will be discussed in more detail in the following section.

\

\section{Discussion of Realistic Setups}\label{sec:simulation}
In this section we analyze the suitability of atoms confined in realistic dipole traps to encode GKP states.
Two types of optical dipole traps are commonly used for experiments with cold atoms: optical tweezers and optical lattices.
The potentials in these platforms are in general not purely harmonic. Therefore, we study the oscillator parameters for these setups analytically and we perform an exemplary state preparation simulation with realistic potentials.
We anticipate here that we find that for a minimal anharmonicity the ratio of the oscillator frequency $\omega_z$ and trap depth must be as small as possible, which can be understood as a high capacity for bosonic excitations.
In order to reduce the coupling between the coding mode $z$ and the spectator modes $x$ and $y$, a high mismatch of oscillator frequencies $\omega_z \ll \omega_{x,y}$ is beneficial.

\subsection{GKP States in an Anharmonic Oscillator} \label{sec:realistic potential shapes}

\begin{figure*}
    \centering
    \includegraphics[width=0.8\linewidth]{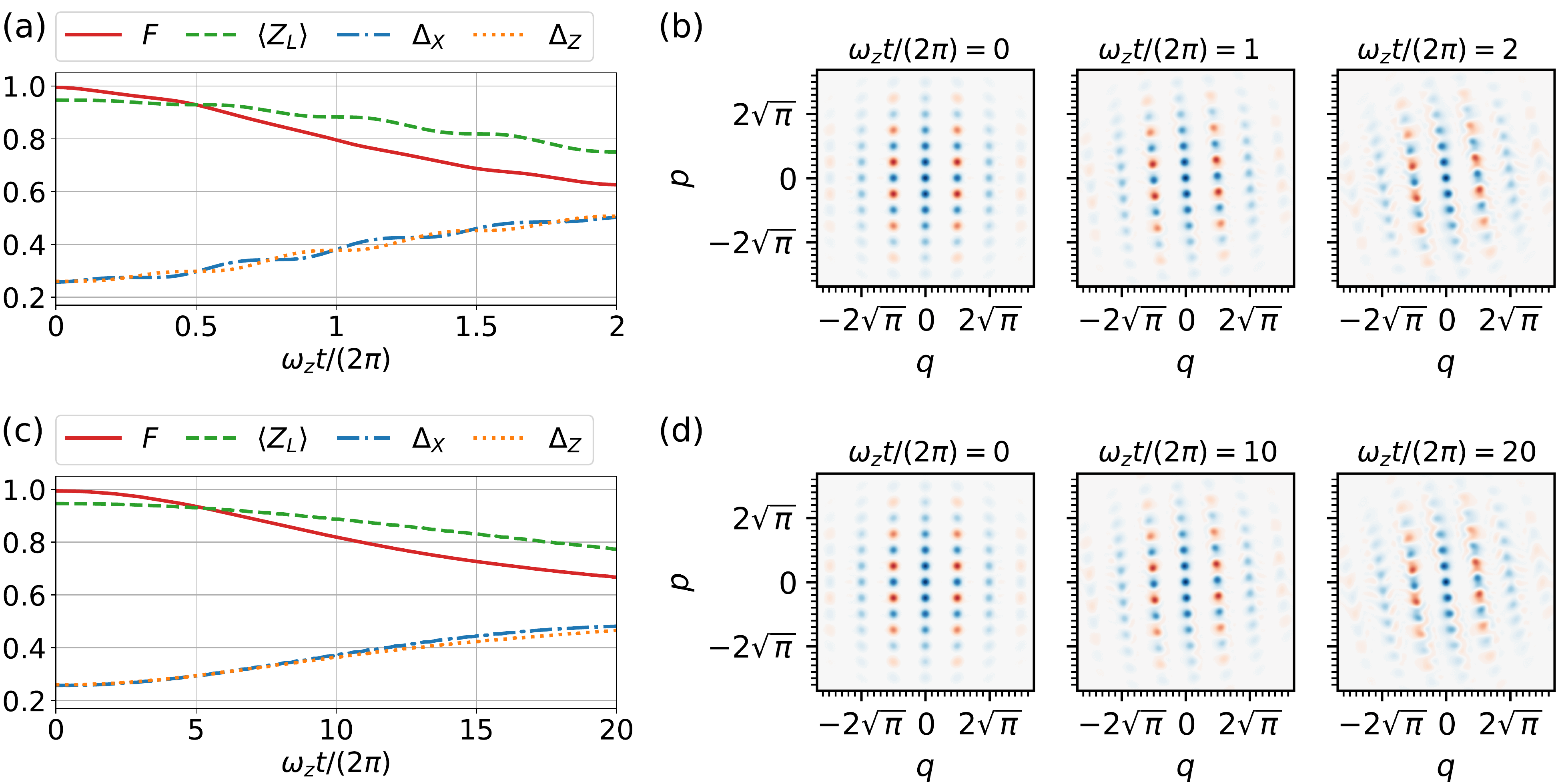}
    \caption{\textbf{Evolution of a finite GKP state in realistic trap potentials.} The figure shows how a finite GKP state $\ket{0_L^{\Delta}}$ with $\Delta = 0.25$ evolves in trap potentials corresponding to: (a,b) an exemplary optical tweezer setup (Table~\ref{tab:tweezer}, Example~I) and (c,d) an exemplary optical lattice setup (Table~\ref{tab:lattice}, Example~II). Panels (a) and (c) display the evolution of the fidelity $F$ w.r.t.~the initial state, the expected logical operator $\langle \mathcal{Z}_L \rangle$ and the effective squeezings $\Delta_{X,Z}$. Panels (b) and (d) show the respective Wigner functions at selected points in time. We define the finite GKP survival time $\tau_{\Delta}$ as the time it takes for the fidelity $F$ to decay to a value of $2/3$. The exemplary tweezer therefore exhibits $\tau_{0.25} \approx 2$ oscillator cycles, while for the exemplary lattice setup we find $\tau_{0.25} \approx 20$ oscillator cycles.}
    \label{fig:finite_GKP_survival_time}
\end{figure*}

Given the potential of a one-dimensional Gaussian laser beam as in Eq.~\eqref{eq:intensity taylor}, the harmonic oscillator is derived by discarding terms of order $\mathcal{O}(z^4)$.
We now consider a multi-dimensional oscillator, which includes couplings between modes of the form $\hat{q}_z^2\hat{q}_x^2$ and also contains anharmonic contributions proportional to $\hat{q}_i^4$.
A two-dimensional anharmonic oscillator Hamiltonian thus reads
\begin{equation} \label{eq:Hamiltonian with eta and epsilon}
    \frac{\hat{H}_\text{anh}}{\hbar \omega_z} = \hat{a}_z^\dagger \hat{a}_z + \frac{\omega_x}{\omega_z}\hat{a}_x^\dagger \hat{a}_x - \eta_z \hat{q}_z^4 - \eta_x \frac{\omega_x}{\omega_z} \hat{q}_x^4 - \varepsilon_{zx}\hat{q}_x^2\hat{q}_z^2 .
\end{equation}
We call the dimensionless coefficients $\eta_i$ anharmonicities and refer to $\varepsilon_{zx}$ as coupling to the mode $x$. For a complete three-dimensional picture, terms with $x \rightarrow y$ must be included as well.
A non-zero anharmonicity $\eta_z$ causes the rotation speed in phase space to decrease with the distance from the origin.
This limits the lifetime and the maximal size of encoded GKP states, which we discuss in more detail below.

The effect of the coupling cannot be treated independently from the behavior inside the unused spectator modes $x$ and $y$. A multidimensional harmonic oscillator that includes couplings proportional to $\hat{q}_i^2\hat{q}_j^2$ is known as Pullen--Edmonds--Hamiltonian~\cite{Pullen:1981}. In contrast to the anharmonicity $\eta_z$, the coupling will not entirely be interpreted as noise, but it will be partially absorbed into a redefinition of the oscillator frequency.
The anharmonic Hamiltonian from Eq.~\eqref{eq:Hamiltonian with eta and epsilon} can be rewritten as follows:
\begin{widetext}
\begin{equation} \label{eq:anharmonic makes elliptic}
    \frac{\hat{H}_\text{anh}}{\hbar \omega_z} = \; \frac{\hat{p}_z^2+\hat{q}_z^2(1-\varepsilon_{zx})}{2}  + \frac{\omega_x}{\omega_z} \hat{a}_x^\dagger \hat{a}_x - \eta_z \hat{q}_z^4 - \eta_x \frac{\omega_x}{\omega_z} \hat{q}_x^4
    - \frac{\varepsilon_{zx}}{4} \left(\hat{a}^{\dagger2}_x + 2\hat
    {a}_x^\dagger \hat{a}_x + \hat{a}^2_x\right)\left(\hat{a}^{\dagger2}_z + 2\hat{a}_z^\dagger \hat{a}_z + \hat{a}^2_z + 1\right) + \mathrm{const} .
\end{equation}
\end{widetext}
This can be interpreted as a redefined oscillator with $\omega_z \mapsto (1-\varepsilon_{zx})^{1/2}\omega_z$ and $\hat{q}_z \mapsto (1-\varepsilon_{zx})^{1/4} \hat{q}_z$, $\hat{p}_z \mapsto (1-\varepsilon_{zx})^{-1/4} \hat{p}_z$. 
For $\omega_x/\omega_z \gg 1$, the terms that include $\hat{a}^{2}_x$ or $\hat{a}^{\dagger2}_x$ can be neglected through a rotating-wave approximation. The term proportional to $\hat{n}_x = \hat{a}_x^\dagger \hat{a}_x$ becomes irrelevant if the $x$-mode can be kept in the ground state. Thus, it is beneficial if the oscillator level spacing, i.e.~the oscillator frequency $\omega_x$ is large.

To quantify the suitability of a trap candidate to host a GKP state we perform the following numerical analysis. We consider an initial finite GKP state $\ket{0_L^{\Delta}}$ and let it undergo time dynamics generated by the trap Hamiltonian of interest, while monitoring the fidelity $F$ w.r.t.~the initial state, the expected logical operator $\langle \mathcal{Z}_L \rangle$ and the effective squeezings $\Delta_{X,Z}$. Physically, this corresponds to idling of the finite GKP state placed in a certain trap potential. We consider the time it takes for the fidelity $F$ to decay to a value of $2/3$ as a simple measure for the trap quality and call it \emph{finite GKP survival time} $\tau_\Delta$. This quantity is not meant to be a rigorous coherence time; it rather serves as a benchmark, which allows one to compare the suitability of different trap candidates. In Fig.~\ref{fig:finite_GKP_survival_time} we show the evolution of a finite GKP state $\ket{0_L^{\Delta}}$ with $\Delta = 0.25$ in two different trap potentials that correspond to an exemplary optical tweezer and an exemplary optical lattice setup. We choose $\Delta = 0.25$ (12\,dB of squeezing) since this value of $\Delta$ is realistic to achieve in experiments, while it was shown to be slightly below threshold for concatenating the GKP code with the surface code~\cite{noh2022low}. From panels (a) and (c) we can identify finite GKP survival times of $\tau_{0.25} \approx 2$ oscillator cycles for the exemplary tweezer setup and $\tau_{0.25} \approx 20$ oscillator cycles for the lattice, which we will discuss in more detail in the following subsection.

\begin{table}[]
{\setlength{\extrarowheight}{5pt}%
\begin{tabular}{l|l|>{\columncolor{gray!15}}r|r|r} \hline \hline
    \multicolumn{2}{c|}{\bf{Optical}} & \underline{Example I} & \underline{Ex. I.b} & \underline{Ex. I.c} \\ 
    \multicolumn{2}{c|}{\bf{Tweezer}} & $w_0=\SI{1}{\micro\metre}$ & $\SI{1.5}{\micro\metre}$ & $\SI{1.5}{\micro\metre}$ \\
    \multicolumn{2}{c|}{} & $U_0/k_B=\SI{1.5}{\milli\kelvin}$ & $\SI{667}{\micro\kelvin}$ & $\SI{1.5}{\milli\kelvin}$ \\ [0.1cm]\hline
$\omega_z$  & $ \frac{\sqrt{2} \lambda}{\pi w_0^2} \sqrt{\frac{U_0}{m}}$ & $2\pi \cdot \SI{28}{\kilo\hertz}$ &$2\pi \cdot \SI{8}{\kilo\hertz}$ &$2\pi \cdot \SI{12}{\kilo\hertz}$ \\[0.2cm]
$\omega_{x,y}$ & $ \frac{2}{w_0} \sqrt{\frac{U_0}{m}}$ & $2\pi \cdot \SI{120}{\kilo\hertz}$ & $2\pi \cdot \SI{53}{\kilo\hertz}$ &$2\pi \cdot \SI{80}{\kilo\hertz}$ \\[0.2cm]
$\eta_z$ & $\frac{1}{4}\frac{\hbar\omega_z}{U_0} $ & $2.2\cdot 10^{-4}$ & $1.5\cdot 10^{-4}$ & $1.0\cdot 10^{-4}$ \\[0.2cm]
$\varepsilon_{zx,zy}$ & $ \frac{1}{2}\frac{\hbar\omega_{x,y}}{U_0}$ & $1.9\cdot 10^{-3}$ & $ 1.9\cdot 10^{-3}$ & $1.3\cdot 10^{-3}$ \\ [0.1cm]\hline \hline
\end{tabular}}
\caption{\textbf{Oscillator parameters for an optical tweezer.} The table shows how the oscillator frequencies $\omega_{x,y,z}$, anharmonicity $\eta_{z}$, and couplings $\varepsilon_{zx,zy}$ depend on the trap depth $U_0$, the wavelength of the dipole trap laser $\lambda$, the beam waist $w_0$, and the mass $m$ of the trapped atom.
The GKP state shall be encoded in the $z$-mode with anharmonicity $\eta_z$, which couples with coefficients $\varepsilon_{zx,zy}$ to modes of frequency $\omega_{x,y}$, as shown in Eq.~\eqref{eq:Hamiltonian with eta and epsilon}. The anharmonicities and couplings can be interpreted as an inverse `boson capacity', i.e. single excitation energy $\hbar\omega$ divided by the trap depth $U_0$.
Three exemplary setups are shown for $^{88}\mathrm{Sr}$ atoms trapped in $\lambda=\SI{1040}{\nano\metre}$ tweezers. Example~I describes a tightly focused tweezer~\cite{lorenz2021Raman,manetsch2024tweezerarray6100}. The trap depth in Ex.~I.b is chosen such that a single tweezer requires the same optical power $P \propto U_0 w_0^2$ as in Ex.~I. The required optical power per tweezer in Ex.~I.c is $2.25$ times larger as compared to the other cases.}
\label{tab:tweezer}
\end{table}

\begin{table}[]
{\setlength{\extrarowheight}{5pt}%
\begin{tabular}{l|l|>{\columncolor{gray!15}}r|r|r} \hline \hline
    \multicolumn{2}{c|}{\bf{Optical}} & \underline{Example II} & \underline{Ex. II.b} & \underline{Ex. II.c} \\ 
    \multicolumn{2}{c|}{\bf{Lattice}} & $w_0=\SI{20}{\micro\metre}$ & $\SI{10}{\micro\metre}$ & $\SI{10}{\micro\metre}$ \\
    \multicolumn{2}{c|}{\bf{}} & $U_0/k_B = \SI{1.5}{\milli\kelvin}$ & $\SI{6}{\milli\kelvin}$ & $\SI{1.5}{\milli\kelvin}$ \\ [0.1cm]\hline
$\omega_z$  & $ \frac{2}{w_0} \sqrt{\frac{U_0}{m}}$ & $2\pi \cdot \SI{6}{\kilo\hertz}$ & $2\pi \cdot \SI{24}{\kilo\hertz}$ & $2\pi \cdot \SI{12}{\kilo\hertz}$ \\[0.2cm]
$\omega_{x,y}$ & $\frac{2 \pi}{\lambda} \sqrt{\frac{U_0}{m}}$ & $2\pi \cdot \SI{362}{\kilo\hertz}$ & $2\pi \cdot \SI{725}{\kilo\hertz}$ & $2\pi \cdot \SI{362}{\kilo\hertz}$ \\[0.2cm]
$\eta_z$ & $\frac{1}{8}\frac{\hbar\omega_z}{U_0} $ & $2.4\cdot 10^{-5}$ & $2.4\cdot 10^{-5}$ & $4.8\cdot 10^{-5}$ \\[0.2cm]
$\varepsilon_{zx,zy}$ & $ \frac{1}{4}\frac{\hbar\omega_{x,y}}{U_0}$ & $2.9\cdot 10^{-3}$ & $ 1.4\cdot 10^{-3}$ & $2.9\cdot 10^{-3}$ \\ [0.1cm]\hline \hline
\end{tabular}}
\caption{\textbf{Oscillator parameters for a 2D optical lattice.} The table shows how the oscillator frequencies $\omega_{x,y,z}$, anharmonicity $\eta_{z}$, and couplings $\varepsilon_{zx,zy}$ depend on the trap depth $U_0$, the wavelength of the dipole trap laser $\lambda$, the beam width $w_0$, and the mass $m$ of the trapped atom.
The GKP state shall be encoded in the $z$-mode.
Three exemplary setups are shown for $^{88}\mathrm{Sr}$ atoms trapped in a $\lambda=\SI{1040}{\nano\metre}$ optical lattice.
Example~II has recently been realized in Ref.~\cite{Tao:2024}, demonstrating a lattice with more than 20,000 sites. The optical power scales as $P \propto U_0 w_0^2$, thus Ex.~II.b requires the same optical power, whereas Ex.~II.c requires less. Note, however, that the size of the lattice array depends on the beam width and therefore Examples~II.b and II.c provide fewer lattice sites than Ex.~II.}
\label{tab:lattice}
\end{table}

\subsection{Comparison of Optical Tweezers and Optical Lattices}
\label{sec:tweezer vs lattice}

An optical tweezer is a highly focused laser beam, where the beam waist at the focal point approaches the diffraction limit given by the wavelength of the light.
This causes the single trap at its focal point to have similar sizes in all three dimensions and thus three modes with similar frequencies, as illustrated in Fig.~\ref{fig:title figure}(b). The full potential follows a Gaussian profile in the radial directions $x$ and $y$ and a Lorentzian profile in the axial direction $z$~\cite{saleh_teich}. For the Gaussian tweezer, the beam waist~$w_0$ and the Rayleigh range~$z_R$ determining the trap size in axial direction are interdependent via $z_R = \pi w_0^2/\lambda$.
An optical lattice, on the other hand, is formed by retroreflected or folded interfering laser beams with transverse waists significantly larger than the wavelength. The trap potential along the beam is determined by interference and thus, the resulting trap spacing is of the order of $\lambda/2$, depending on the exact implementation (see Fig.~\ref{fig:title figure}(b)).
In a 2D optical lattice, the potential in $x$ and $y$ is created by beam interference, whereas the confinement in the $z$-direction is determined by the beam width.
The potential is described by a cosine in $x$ and $y$, while it follows a Gaussian profile in the perpendicular $z$-direction.
Details on the potential shapes are provided in Appendix~\ref{appendix: Tweezer and Lattice Potentials}.
The potentials of an optical tweezer and a 2D optical lattice can be expanded to determine the oscillator frequencies, anharmonicities and couplings analytically.
These quantities follow the equations shown in Tables~\ref{tab:tweezer} and \ref{tab:lattice}.

Considering Table~\ref{tab:tweezer} for an optical tweezer, we see that the ratio of radial and axial trap frequencies is given by $\omega_x / \omega_z = \sqrt{2} \pi w_0/\lambda$. The anharmonicity $\eta_z$ and the couplings $\varepsilon_{zx,zy}$ can be interpreted as an inverse `boson capacity', i.e. single excitation energy $\hbar\omega$ divided by the trap depth $U_0$.
Moreover, the Table displays concrete numbers for three exemplary tweezer setups. To be specific, we examine $^{88}\mathrm{Sr}$ atoms with mass $m = \SI{87.9}{u}$ and consider a trap laser with wavelength $\lambda = \SI{1040}{\nano \metre}$. More details on the choice of ancilla qubit and therefore the wavelength of the laser will follow in the next subsection.
Two parameters that one can vary in an optical tweezer experiment are the beam waist~$w_0$ and the trap depth~$U_0$. The optical power per beam, however, scales as $P \propto U_0 w_0^2$. Therefore, increasing the beam waist at constant trap depth requires increasing optical power per individual tweezer, which in turn limits the size of the tweezer array. The highlighted setup Example~I ($w_0 = \SI{1}{\micro \metre}$, $U_0/k_B = \SI{1.5}{\milli \kelvin}$) corresponds to a tightly focused tweezer~\cite{lorenz2021Raman,manetsch2024tweezerarray6100}. The time evolution of a finite GKP state in this trap is shown in Fig.~\ref{fig:finite_GKP_survival_time}(a,b), indicating a finite GKP survival time $\tau_{0.25} \approx 2$ oscillator cycles. The trap anharmonicity can be reduced by increasing the beam waist $w_0$ at constant optical power, corresponding to Ex.~I.b in the Table. However, we see that this comes at the expense of a shallower trap and smaller trap frequencies of the $z$-mode as well as the spectator modes. Increasing the waist $w_0$ at constant trap depth reduces the anharmonicity even further, as can be seen from Ex.~I.c; however, the higher optical power requirements per individual tweezer limit the number of traps that can be produced for the total array.

Table~\ref{tab:lattice} displays the oscillator characteristics for an anisotropic two-dimensional optical lattice setup. A crucial difference to the tweezer array is that a single beam with a large vertical waist $w_0$ is used to create the entire lattice. While this large waist sets the trap frequencies along the vertical $z$-direction, the trap frequencies $\omega_x$ and $\omega_y$ are entirely determined by the trap depth and wavelength of the trapping laser. The frequency ratio given by $\omega_x / \omega_z = \pi w_0/\lambda$ is therefore larger for the lattice. Again, the anharmonicity $\eta_z$ and the couplings $\varepsilon_{zx,zy}$ can be interpreted as an inverse boson capacity. As for the tweezer, three exemplary setups are shown for $^{88}\mathrm{Sr}$ atoms trapped by a laser with wavelength $\lambda = \SI{1040}{\nano \metre}$. The parameters of Example~II correspond precisely to a recent experiment realized in Ref.~\cite{Tao:2024}. In Fig.~\ref{fig:finite_GKP_survival_time}(c,d) we show the time evolution of a GKP state in this trap potential from which we deduce a finite GKP survival time $\tau_{0.25} \approx 20$ oscillator cycles. Example~II.b considers a smaller beam waist at constant optical power. It comes with the disadvantage of a smaller number of trap sites but provides larger absolute oscillator frequencies. Example~II.c shows the parameters for a smaller vertical waist at constant trap depth. This setup requires less optical power but exhibits an increased trap anharmonicity. We note that the power overhead could be used to enlarge the lattice array.

Comparing the suitability of optical tweezers and optical lattices to host GKP states, we first examine the number of trapped atoms which can be operated in these setups at comparable optical powers. The lattice setup Ex.~II. has demonstrated the creation of an optical lattice array with more than 20,000 sites, which can be operated continuously with more than 1,000 sorted individual atoms~\cite{Tao:2024, Gyger:2024}. In a tweezer array the total optical power is divided evenly among the tweezers. This means that larger arrays can be assembled if the power requirements per individual tweezer are small.
For the same total power that allows one to realize more than 10,000 lattice traps in Ex.~II~\cite{Tao:2024}, at most a few hundred tweezers can be produced for the setups in Ex.~I and I.b for Strontium under realistic assumptions.
As discussed in the previous subsection, a large frequency mismatch between the coding mode and the spectator modes is beneficial to minimize detrimental effects that result from the coupling between modes, which holds particularly true for the optical lattices.
Moreover, we find that the optical lattice setups obey smaller trap anharmonicities than the tweezer examples.
When it comes to idling of GKP states, i.e.~preserving encoded information for as long as possible, a small anharmonicity is crucial and we therefore conclude that a typical optical lattice setup is advantageous as compared to an optical tweezer array in this regard.
Eventually, however, a truly flexible setup would be a static optical lattice augmented with dynamical optical tweezers. We envision an optical lattice that is sparsely filled with atoms, each encoding a GKP qubit. Optical tweezers can then be employed to perform operations such as displacements on individual atoms without crosstalk.

\subsection{Exemplary GKP State Preparation in an Optical Lattice}
\label{sec:final preparation}

Finally, we discuss a concrete implementation example and we numerically simulate the preparation of a finite GKP state by implementing the scheme shown in Fig.~\ref{fig:GKP sharpen circuit} utilizing realistic tweezer and lattice potentials. As a specific example, and based on the preceding discussion, we consider atoms trapped in an optical lattice.

Requirements for the electronic ancilla qubit are the existence of magic and tune-out wavelengths, and sufficiently fast single-qubit gates and qubit reset. Here we briefly discuss the necessity of fast single-qubit gates. The state preparation and error correction protocol shown in Fig.~\ref{fig:GKP sharpen circuit} requires the repeated application of single-qubit gates on the ancilla. It is therefore crucial that single-qubit gates can be executed on a time scale that is very short as compared to the finite GKP survival time. For a single-qubit Rabi frequency $\Omega$ one thus requires that $\Omega \tau_{\Delta} / (2\pi) \gg 1$.
The exemplary setup we consider exhibits a finite GKP survival time $\tau_{0.25} \approx 20$ oscillator cycles, where the oscillator frequency is $\omega_z = 2 \pi \times \SI{6}{\kilo \hertz}$.
A suitable ancilla qubit candidate is therefore the electronic fine-structure qubit in $^{88}\mathrm{Sr}$ formed by the two metastable states $^3\!P_0$ and $^3\!P_2$. Rabi frequencies of the order of $2\pi \times\SI{100}{\kilo \hertz}$ have been demonstrated  for this qubit encoding~\cite{Unnikrishnan:2024, Pucher:2024}, such that $\Omega \tau_{0.25} / (2 \pi) > 300$.
Another prominent qubit in $^{88}\mathrm{Sr}$ is the optical clock qubit formed by the states $^1\!S_0$ and $^3\!P_0$~\cite{Heinz:2020, Young_2020_half, Schine_2022_long}. In recent experimental works, single-qubit Rabi oscillations with Rabi frequencies $\Omega \approx 2\pi \times \SI{1}{\kilo\hertz}$ have been achieved~\cite{finkelstein2024universal, cao2024multiqubit}. The duration of a single-qubit gate in the optical clock qubit is therefore of the same order of magnitude as the oscillator period and $\Omega \tau_{0.25} / (2 \pi) \approx 3$. Utilizing the clock qubit would therefore prolong every cycle of the state preparation procedure by several oscillator periods. This increases the time for which the atoms are exposed to the anharmonic potential deteriorating the state steadily. However, if the quantity $\Omega \tau_{\Delta}$ can be enhanced, e.g.~by increasing the single-qubit Rabi frequency working with three-photon coupling~\cite{he2024coherent,carman2024collinear,ammenwerth2024realization}, or by using a trap that exhibits a larger finite GKP survival time, the clock qubit could become equally suitable. Here we continue our discussions considering the fine-structure qubit in $^{88}\mathrm{Sr}$ and therefore consider a laser wavelength $\lambda = \SI{1040}{\nano \metre}$, which can trap both states of the fine-structure qubit in approximately the same potential, such that the qubit decouples from the bosonic modes (magic wavelength). Also fast reset of the fine-structure qubit is possible~\cite{ammenwerth2024realization}. However, to minimize heating caused by photon recoil during the reset, larger trap frequencies allowing for smaller Lamb--Dicke parameters are beneficial.

We now proceed with the simulation of the preparation of GKP states in a realistic optical lattice, utilizing an optical tweezer to implement the displacements.
For the trap Hamiltonian, the kinetic energies and the full three-dimensional lattice potential (see Appendix~\ref{appendix: Tweezer and Lattice Potentials}) are evaluated via matrix exponentiation in the Fock basis of the coupling-corrected harmonic oscillators. As lattice parameters, the aforementioned parameters from Table~\ref{tab:lattice}, Example~II are chosen, resulting in an oscillator with frequency $\omega_z \approx 2\pi \times \SI{6}{\kilo\hertz}$ and anharmonicity $\eta_{z} \approx 2.4 \times 10^{-5}$.

\begin{figure}[t]
\centering
\includegraphics[width=0.99\linewidth]{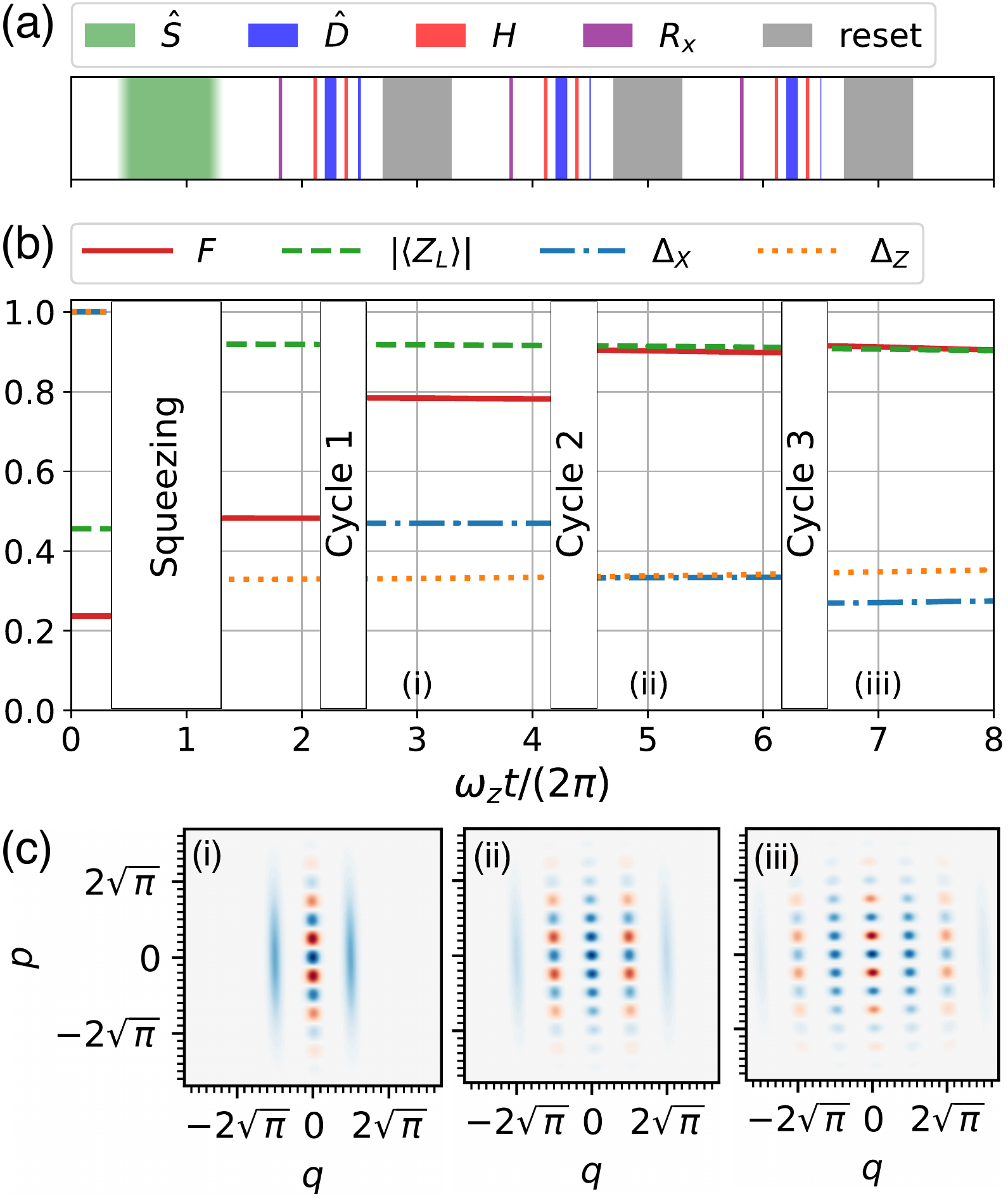}
    \caption{\textbf{Simulation of GKP state preparation in an optical lattice.} (a)~Pulse scheme for the preparation of a GKP state, see Fig.~\ref{fig:GKP sharpen circuit}. (b)~Effective squeezings $\Delta_{X,Z}$, expected logical operator $|\langle \mathcal{Z}_L \rangle|$ and fidelity $F$ w.r.t.~a reference GKP state in the course of the protocol. Note that the protocol prepares a logical state of type $\ket{0_L}$ after even numbers of preparation cycles, while the state is of type $\ket{1_L}$ after odd numbers of cycles, as explained in Sec.~\ref{sec:II.C GKP state prep}. The reference state for the fidelity measure therefore alternates between $\ket{0_L^{0.25}}$ and $\ket{1_L^{0.25}}$ with every cycle. All quantities are obtained from a state simulation discussed in the main text. The procedure starts in the vacuum state with $\Delta_X=\Delta_Z=1$. The squeezing pulse yields $\Delta_Z\approx0.32 \approx 1/\Delta_X $. The displacement cycles reduce $\Delta_X$ and increase the fidelity $F$ stepwise, while the oscillator anharmonicity and the coupling between modes continuously degrade the state. (c)~Wigner functions of the prepared states after each preparation cycle, i.e. after oscillator cycles 3, 5 and 7. The effect of the anharmonicity is visible as a non-ideal rotation at the edge of the state and the effect of the coupling is visible as a slight smearing on the outer dots of the state.}
    \label{fig:Pulse scheme}
\end{figure}

The squeezing procedure depicted in Fig.~\ref{fig:tweezer squeezing demonstration} is performed with a temporary potential decrease to $U_0/10$, switching to an elliptic rotation with $\omega' = \omega_z/\sqrt{10} \approx 2\pi \times \SI{1.9}{\kilo\hertz}$. The necessary duration in that oscillator then is $2\pi/(4\omega') \approx \SI{132}{\micro\second}$, which corresponds to roughly $80\%$ of a normal oscillation cycle time of $\approx \SI{167}{\micro\second}$.
To keep the $x$- and $y$-modes in the ground state, the switch to the lower intensity is not done instantaneously, but over a ramping duration of $\SI{20}{\micro\second}$. This causes the dynamics of the state in the $z$-mode to deviate slightly from the ideal dynamics, but appears mostly as an adiabatic change to the $x$- and $y$-modes, which temporarily slow down from an oscillator period of $\approx \SI{2.76}{\micro\second}$ to $\approx \SI{8.74}{\micro\second}$.
In Appendix~\ref{appendix: squeezing, displacement} we analyze the squeezing procedure by means of a fast decrease of the trap depth in more detail.
The simulation indicates that a ground state population $\bra{0_x0_y}\text{Tr}_z \left(\rho\right) \ket{0_x0_y} > 98\%$ in the $x$- and $y$-modes can be sustained with these parameters. Applying pulse profiles that are optimized using techniques from control theory can potentially improve this even further~\cite{Werschnik_2007, grochowski2024}.
Excitations in the $x$- or $y$-mode lead to a further slow-down of the oscillator in $z$, similar to the shift in Eq.~\eqref{eq:anharmonic makes elliptic}.
Note that later error correction cycles applied on the GKP state encoded in the $z$-mode do not remove excitations in the spectator modes.
As a worst-case estimate, the infidelity from excitations in $x$ and $y$ can be considered as lost, decreasing the fidelity once in the beginning.
In our simulations we consider a Hilbert space with dimension $18\times18\times36$, where $n_z=36$ is the highest considered excitation of the GKP mode and $n_x=n_y=18$ is the cut-off for the $x$- and $y$-modes. This results in a squeezed state with $\Delta_Z < 0.35$ and $\Delta_X > 1$ (see Fig.~\ref{fig:Pulse scheme}). After the squeezing procedure, the simulation changes to a mixed state simulation in a Hilbert space with dimension $4\times4\times36$, which is enough to incorporate first-order effects of the couplings $\hat{a}_x^{\dagger 2}$, $\hat{a}_x^\dagger \hat{a}_x$ and $\hat{a}_x^2$.

The displacement sequence is performed according to the deterministic scheme shown in Fig.~\ref{fig:GKP sharpen circuit} for three repetitions with $\epsilon_{k}=0$ and values $\delta_k$ which produce a minimal effective squeezing $\Delta_X$ under ideal displacements, discussed in Sec.~\ref{sec:II.C GKP state prep}. To include all orders of anharmonicity and coupling, the displacements have not been implemented as ideal unitaries, but by addition of tune-out potentials from adequately placed one-dimensional Gaussian laser beams of width $w_1=\SI{20}{\micro\metre}$.
For this proof-of-principle simulation we do not take tweezers addressed to individual atoms. We rather consider a single large tweezer that is used to illuminate all atoms simultaneously in order to prepare each atom in a GKP state. Pictorially, for an optical lattice array placed in the $x$-$y$-plane and motional states encoded in the $z$-mode of trapped atoms, we consider a large tweezer whose axis is oriented perpendicular to the z-axis and slightly shifted out of the x-y-plane.
For simplicity, we place two such displacement lasers, one with a $\ket{0}$-tune-out wavelength and one with $\ket{1}$-tune-out wavelength. Their potential peak is chosen to be $U_1/\hbar= 2\pi \times \SI{2}{\mega\hertz}$, which is around $7\%$ of the trap depth.
The force $f$ of these beams then moves the effective trap center by $|\alpha_d| \approx 2.81$ in opposing directions conditioned on the ancilla state. A stabilizer displacement $\hat{D}(\pm\sqrt{\pi})$ then requires a pulse duration $\omega t /(2\pi) \approx  0.1020$ according to Eq.~\eqref{eq:Momentum kick} (see the longest displacement pulse in Fig.~\ref{fig:Pulse scheme}(a)).
To achieve the targeted $\delta_1 = 1$ in the first round, pulse durations $\omega t / (2 \pi) \approx 0.0251$ are needed. The second round needs for $\delta_2 = 0.5$ approximately half of that with $\omega t / (2 \pi) \approx 0.0125$ and the third one for $\delta_3 = 0.303$ even less with $\omega t / (2 \pi) \approx 0.0076$.
The single-qubit operations of the electronic ancilla are modelled as ideal rotations with a Rabi frequency $\Omega = 2\pi \times \SI{100}{\kilo\hertz}$, which is faster than the bosonic mode by a factor of approximately $20$.

As seen in Fig.~\ref{fig:Pulse scheme}(b), the first displacement cycle reduces $\Delta_X$ from $3.33$ to roughly $0.47$. The second and third cycle lower it further to $\Delta_X \approx 0.33$ and $\Delta_X \approx 0.27$, respectively.
The fidelity w.r.t. to the corresponding ``target'' GKP state with $\Delta = 0.25$ increases with every preparation cycle.
The effective squeezing $\Delta_Z$ and the expected logical operator $|\langle \mathcal{Z}_L \rangle|$ are mainly unaffected by the displacement procedures. However, the anharmonicity and coupling of the trap beam and displacement beam slowly degrade all of these quantities over time. This effect is still quite moderate, indicating that the effective squeezings can still be decreased further with current levels of anharmonicity in a lattice.

Imperfect ground state initialization due to finite cooling will lead to initial effective squeezings $\Delta_X, \Delta_Z > 1$.
In an experiment, recoil heating and trap noise, in particular relative intensity noise~\cite{grimm_2000_optical, savard1997laser, Gehm:1998}, will continuously deteriorate the state as well. For simplicity, they are not part of this simulation since they are not expected to dominate the behavior during the preparation.
To get an idea of the time scale on which heating has to be taken into account, a rough calculation can be performed. Considering the lattice setup in Ref.~\cite{Tao:2024}, the lifetime of atoms in the traps is of the order of several seconds. Taking a lifetime of $\SI{10}{\second}$, trap frequencies in the $x$- and $y$-modes of $\SI{300}{\kilo \hertz}$ and a trap depth of $\SI{30}{\mega \hertz}$ (corresponding to $\SI{1.5}{\milli \kelvin}$), we obtain about 100 confined levels. This implies the increase of the temperature by one motional quantum every $\SI{100}{\milli \second}$, corresponding to 600 oscillator cycles of the $z$-mode with frequency $\SI{6}{\kilo \hertz}$.
This very rough estimate of the heating time scale is orders of magnitude larger than the time scale on which effects of the trap anharmonicity become visible. Note that in this estimation we use the trap frequency of the spectator modes because heating rates increase with trap frequencies in optical traps~\cite{savard1997laser} and are thus dominated by the mode with the largest trap frequency.

In Fig.~\ref{fig:Pulse scheme}(c) we show the Wigner functions of the prepared states after each preparation cycle. The anharmonic contributions appear as non-ideal rotations towards the edges of the state. However, overall the final state closely resembles the desired state from Fig.~\ref{fig:GKP sharpen circuit}, which demonstrates the successful state preparation.

\section{Conclusion and Outlook}\label{sec:conclusion}

In this paper we have shown how GKP states can be prepared deterministically in the motional mode of an atom confined in an optical dipole trap.
We suggest to use a deterministic preparation scheme, adapted from error correction of GKP states, and numerically identify optimal corrective displacement distances $\delta_k$ for $k$ preparation cycles.
We outline how squeezing and conditional displacements can be realized on atoms confined in optical dipole traps and argue that a sparsely filled optical lattice is a more promising candidate than a tweezer array due to its lower anharmonicity and better scalability under the constraints given by the goal to create high-fidelity GKP states.
To realize squeezing one can temporarily decrease the trap depth to around $10\%$ of its default value. We find that with an adequate ramping speed the squeezing can be limited to the mode in which the GKP state is encoded, while leaving the other two vibrational modes close to the ground state.
For a conditional displacement, an additional optical tweezer beam operating at the tune-out wavelength can be utilized. Here we show that typical beam intensities are more than sufficient for rapid displacements of the required distances.
All in all, we demonstrate that a state with effective squeezings $\Delta_X\approx\Delta_Z \approx 0.3$ can be prepared in five oscillator cycles or less and explain how error correction can be realized using the same protocol with different parameters.

As a continuation beyond the scope of this paper, one can analyze the effect of atom heating due to photon recoil and trap noise and their interplay with the error correction procedure.
Furthermore, the effect of the finite GKP size $\Delta$ and other experimental parameters on the GKP survival time can be investigated. Another interesting question is whether the trap shape of a tweezer can be optimized using spatial light modulators in order to minimize the anharmonicity.
Moreover, one could investigate how to realize logical gate protocols~\cite{liu2024hybrid,rojkov2024}, e.g. how Rydberg-Rydberg interactions between atoms could be used to implement entangling gates between GKP qubits.
At the same time, we judge the preparation and correction protocol to be ready for experimental realizations in state-of-the-art optical lattice setups.

\begin{acknowledgements}
We thank Andrea Alberti, Eran Reches and Wojciech Adamczyk for fruitful discussions.
This research is part of the Munich Quantum Valley (K-3, K-8), which is supported by the Bavarian state government with funds from the Hightech Agenda Bayern Plus.
We additionally acknowledge support by the German Federal Ministry of Education and Research (BMBF) project MUNIQC-ATOMS (Grant No. 13N16070).
M.M., D.F.L. and L.H.B. gratefully acknowledge funding by the German Research Foundation (DFG) under Germany’s Excellence Strategy ‘Cluster of Excellence Matter and Light for Quantum Computing (ML4Q) EXC 2004/1’ 390534769,
by the BMBF via the VDI within the project IQuAn (Grant No. 13N15677),
and by the European Research Council (ERC) Starting Grant QNets through Grant No. 804247.
J.Z. acknowledges support from the BMBF through the program ``Quantum technologies -- From basic research to market'' (Grant No. 13N16265, SNAQC).
J.Z. is a co-founder of planqc.
\end{acknowledgements}

\FloatBarrier
\clearpage
\newpage
\newpage

\appendix
\onecolumngrid

\section{Finite GKP States} \label{appendix: finite GKP}

An important premise is that finite GKP states $\ket{\psi_L^{\Delta}} = 2\sqrt{\pi}\Delta e^{-\Delta^2\hat{a}^\dagger \hat{a}}\ket{\psi_L}$ can be interpreted as superpositions of squeezed states. Schematically, the $\hat{p}^2$-part of $\hat{a}^\dagger \hat{a}= (\hat{p}^2+\hat{q}^2-1)/2$ transforms the position-eigenstates into squeezed states and the $\hat{q}^2$-part reduces their individual contributions by a Gaussian envelope in $\hat{q}$. Here we provide a derivation of Eq.~\eqref{eq:GKP superposition of squeezed states} in the main text.
\begin{equation} \label{eqApp:GKP superposition of squeezed states long}
    \begin{aligned}
        \ket{0_L^{\Delta}} & = 2\sqrt{\pi}\Delta e^{-\Delta^2\hat{a}^\dagger \hat{a}} \ket{0_L} \\
        & \propto e^{-\frac{\Delta^2}{2}(\hat{p}^2+\hat{q}^2)} \sum_{k=-\infty}^{\infty} \ket{q\!=\!2k\sqrt{\pi}} \\
        & = \sum_{k=-\infty}^{\infty} e^{-2\pi \Delta^2 k^2}e^{-\frac{\Delta^2}{2}\hat{p}^2}\ket{q\!=\!2k\sqrt{\pi}} + \mathcal{O}(\Delta^4) \\
        & = \int dp \sum_{k=-\infty}^{\infty} e^{-2\pi \Delta^2 k^2}e^{-\frac{\Delta^2}{2}\hat{p}^2} \ket{p}\!\bra{p} \ket{q\!=\!2k\sqrt{\pi}} + \mathcal{O}(\Delta^4) \\
        & \propto \int dp \sum_{k=-\infty}^{\infty} e^{-2\pi \Delta^2 k^2} e^{-\frac{(\Delta p)^2}{2}} \ket{p} e^{-2ik\sqrt{\pi} p} + \mathcal{O}(\Delta^4) \\
        & = \int dp \sum_{k=-\infty}^{\infty} e^{-2\pi \Delta^2 k^2} e^{-2ik\sqrt{\pi} \hat{p}} \ket{p} e^{-\frac{(\Delta p)^2}{2}} + \mathcal{O}(\Delta^4) \\
        & \propto \int dp \sum_{k=-\infty}^{\infty} e^{-2\pi \Delta^2 k^2} \hat{D}(2k\sqrt{\pi}) \ket{p}\!\bra{p} \hat{S}(-\ln \Delta) \ket{0} + \mathcal{O}(\Delta^4) \\
        & = \sum_{k=-\infty}^{\infty} e^{-2 \pi \Delta^2 k^2} \hat{D}(2k\sqrt{\pi}) \hat{S}(-\ln\Delta)\ket{0} + \mathcal{O}(\Delta^4) .
    \end{aligned}
\end{equation}
In the derivation we used that the momentum space wavefunction of the position-squeezed vacuum is a Gaussian: $\bra{p} \hat{S}(r) \ket{0} \propto e^{-\frac{1}{2}(\frac{p}{e^{r}})^2}$, with $r \in \mathds{R}$~\cite{schumaker_1986_Quantum}.

\section{Prepared States after up to Three Rounds} \label{appendix: prepared states explicitly}

In this Appendix we derive analytically the finite GKP states prepared with the corrective displacement scheme described in Sec.~\ref{sec:II.C GKP state prep} and Fig.~\ref{fig:GKP sharpen circuit}. To do this we make use of the braiding relation stated in Eq.~\eqref{eq:braiding relation}.
As short notation, the corrective displacement $C^{\xi} \coloneq \hat{D}(\xi i\sqrt{\pi}/2)$, the stabilizer $S^{\xi} \coloneq \hat{D}(2\xi \sqrt{\pi})$ and the complex numbers $(-1)^\xi \coloneq e^{\pi i \xi}$ are used. The derivation only needs the braiding relation $C^{\xi}S^{\zeta} = (-1)^{\xi \zeta} S^{\zeta}C^{\xi}$, so it also works with non-square GKP grids. The symbol $\oplus$ is used to denote a mixture of operators in the sense of $(\hat{A} \oplus \hat{B}) (\rho) \coloneq \hat{A} \rho \hat{A}^\dagger + \hat{B} \rho \hat{B}^\dagger$.
One round of stabilizer application and corrective displacement is equivalent to the application of the following channel:
\begin{equation}
    \begin{aligned}
        & C^{\frac{\delta}{2}} \frac{S^{\frac{1}{2}}+iS^{\frac{1}{2}}}{2} \oplus C^{-\frac{\delta}{2}} \frac{S^{\frac{1}{2}}-iS^{\frac{1}{2}}}{2} \\
        = \;& \frac{(-1)^\frac{\delta}{4} S^\frac{1}{2} + i (-1)^\frac{-\delta}{4} S^\frac{-1}{2}}{2} C^\frac{\delta}{2} \oplus \frac{(-1)^\frac{-\delta}{4} S^\frac{1}{2} - i (-1)^\frac{\delta}{4} S^\frac{-1}{2}}{2} C^\frac{-\delta}{2} \\
        \xrightarrow{\delta=1} \; & \underbrace{\frac{S^{\frac{1}{2}}+S^{\frac{-1}{2}}}{2}}_{\text{pair state}} \left((-1)^{\frac{1}{4}}C^\frac{1}{2} \oplus (-1)^{\frac{-1}{4}} C^{\frac{-1}{2}} \right)
    \end{aligned}
\end{equation}
The mixture is caused by the reset of the ancilla qubit: Just before reset, the left summand is entangled to $\ket{0}$ and the right summand to $\ket{1}$. For $\delta = 1$ in the first round the effective stabilizer application is independent of the ancilla state and only the momentum shifts $C$ are mixed. This is why the pre-procedure state must be an approximate eigenstate of this operator.
Combining a second round with a different $\delta$ after $\delta=1$ creates the combined effective channel:
\begin{equation}
    \begin{aligned}
        & \frac{(-1)^\frac{\delta}{4} S^\frac{1}{2} + i (-1)^\frac{-\delta}{4} S^\frac{-1}{2}}{2} C^\frac{\delta}{2} \overbrace{\frac{S^{\frac{1}{2}}+S^{\frac{-1}{2}}}{2}}^{\text{from round 1}}\\
        \oplus \; & \frac{(-1)^\frac{-\delta}{4} S^\frac{1}{2} - i (-1)^\frac{\delta}{4} S^\frac{-1}{2}}{2} C^\frac{-\delta}{2} \overbrace{\frac{S^{\frac{1}{2}}+S^{\frac{-1}{2}}}{2}}^{\text{from round 1}} \\
        =\; & \frac{(-1)^\frac{\delta}{4} S^\frac{1}{2} + i (-1)^\frac{-\delta}{4} S^\frac{-1}{2}}{2} \cdot \frac{(-1)^\frac{\delta}{4} S^\frac{1}{2} +  (-1)^\frac{-\delta}{4} S^\frac{-1}{2}}{2} C^\frac{\delta}{2} \\
        \oplus \; & \frac{(-1)^\frac{-\delta}{4} S^\frac{1}{2} - i (-1)^\frac{\delta}{4} S^\frac{-1}{2}}{2} \cdot \frac{(-1)^\frac{-\delta}{4} S^\frac{1}{2} - (-1)^\frac{\delta}{4} S^\frac{-1}{2}}{2} C^\frac{-\delta}{2} \\
        =\; & \frac{(-1)^\frac{\delta}{2} S^1 + (1+i)  + i(-1)^\frac{-\delta}{2} S^{-1}}{4} C^\frac{\delta}{2} \\
        \oplus \; & \frac{(-1)^\frac{-\delta}{2} S^1 + (1-i)  - i(-1)^\frac{\delta}{2} S^{-1}}{4} C^\frac{-\delta}{2} \\
        \xrightarrow{\delta=\frac{1}{2}} \; & \underbrace{\frac{S+\sqrt{2}+S^{-1}}{4}}_{\text{three positions}} \left( (-1)^\frac{1}{4} C^\frac{1}{4} \oplus (-1)^{\frac{-1}{4}}C^{\frac{-1}{4}} \right)
    \end{aligned}
\end{equation}
For $\delta=\frac{1}{2}$ after $\delta=1$ the effective stabilizer application again decouples from the ancilla qubit. The new mixture of $C$ combines with the one from the first round to a stronger disturbance of the initial state.
The third round can be treated similarly:
\begin{equation}
    \begin{aligned}
        & \frac{(-1)^\frac{\delta}{4} S^\frac{1}{2} + i (-1)^\frac{-\delta}{4} S^\frac{-1}{2}}{2} C^\frac{\delta}{2} \overbrace{\frac{S^{1}+\sqrt{2} + S^{-1}}{4}}^{\text{from round 2}}\\
        \oplus \; & \frac{(-1)^\frac{-\delta}{4} S^\frac{1}{2} - i (-1)^\frac{\delta}{4} S^\frac{-1}{2}}{2} C^\frac{-\delta}{2} \overbrace{\frac{S^{1}+\sqrt{2} + S^{-1}}{4}}^{\text{from round 2}}\\
        = \; & \frac{(-1)^\frac{\delta}{4} S^\frac{1}{2} + i (-1)^\frac{-\delta}{4} S^\frac{-1}{2}}{2} \cdot \frac{(-1)^{\frac{\delta}{2}}S^{1}+\sqrt{2} + (-1)^{\frac{-\delta}{2}}S^{-1}}{4} C^\frac{\delta}{2} \\
        \oplus \; & \frac{(-1)^\frac{-\delta}{4} S^\frac{1}{2} - i (-1)^\frac{\delta}{4} S^\frac{-1}{2}}{2} \cdot \frac{(-1)^{\frac{-\delta}{2}}S^{1}+\sqrt{2} + (-1)^{\frac{\delta}{2}}S^{-1}}{4} C^\frac{-\delta}{2} \\
        = \; & \frac{(-1)^{\frac{3}{4}\delta}S^{\frac{3}{2}}+(i+\sqrt{2})(-1)^{\frac{1}{4}\delta}S^{\frac{1}{2}}+ (1+i\sqrt{2})(-1)^{\frac{-1}{4}\delta}S^{\frac{-1}{2}}+i(-1)^{\frac{-3}{4}\delta}S^{\frac{-3}{2}}}{8} C^{\frac{\delta}{2}} \\
        \oplus \; & \frac{(-1)^{\frac{-3}{4}\delta}S^{\frac{3}{2}}+(-i+\sqrt{2})(-1)^{\frac{-1}{4}\delta}S^{\frac{1}{2}}+ (1-i\sqrt{2})(-1)^{\frac{1}{4}\delta}S^{\frac{-1}{2}}-i(-1)^{\frac{3}{4}\delta}S^{\frac{-3}{2}}}{8} C^{\frac{-\delta}{2}} \\
        \xrightarrow{\delta=\frac{1}{3}} 
        \; & \frac{(-1)^{\frac{1}{4}}S^{\frac{3}{2}}+(i+\sqrt{2})(-1)^{\frac{1}{12}}S^{\frac{1}{2}}+ (1+i\sqrt{2})(-1)^{\frac{-1}{12}}S^{\frac{-1}{2}}+i(-1)^{\frac{-3}{12}}S^{\frac{-3}{2}}}{8} C^{\frac{1}{6}} \\
        \oplus \; & \underbrace{\frac{(-1)^{\frac{-1}{4}}S^{\frac{3}{2}}+(-i+\sqrt{2})(-1)^{\frac{-1}{12}}S^{\frac{1}{2}}+ (1-i\sqrt{2})(-1)^{\frac{1}{12}}S^{\frac{-1}{2}}-i(-1)^{\frac{1}{4}}S^{\frac{-3}{2}}}{8}}_{\text{superposition of four states, mixed with a similar superposition above}} C^{\frac{-1}{6}} \\
    \end{aligned}
\end{equation}
From the third round on the stabilizer application does not decouple from the ancilla for any $\delta$, so another property has to optimized, for example the expected stabilizer value of the prepared state.

\section{Optical Dipole Traps} \label{appendix: optical dipole traps}

Here, for completeness, we review very briefly a general model for optical dipole traps~\cite{phillips_1998_nobel,grimm_2000_optical}, resulting in a potential that is proportional to the local laser light intensity.
To describe the atom-light interaction, a classical wave description of the light is chosen and combined with an electronic two-level system. This two-level system should not be confused with the electronic ancilla qubit. Both $\ket{0}$ and $\ket{1}$ of the ancilla independently couple to different additional states. The two-level system with energy spacing $\hbar \omega_e$ can be transformed into a rotating frame of the light frequency $\omega_L$ such that
after performing a rotating wave approximation the Hamiltonian reads
\begin{equation} \label{eq:Atom in light two level Hamiltonian}
    \hat{H}_{\mathrm{trap}}/\hbar = \frac{\delta}{2} \hat{\sigma}_z +  \frac{\Omega}{2} \hat{\sigma}_x ,
\end{equation}
where $\delta = \omega_e-\omega_L$ is the detuning
and $\Omega = \vec{\mu} \cdot \vec{E}/\hbar$ is the Rabi frequency, which is proportional to the transition dipole moment $\vec{\mu}$ and the electric field amplitude $\vec{E}$, and which is assumed to be real and positive here~\cite{Foot:2005}.
For $\delta \gg \Omega$, which is called the far-detuned regime, one can treat the $\hat{\sigma}_x$ part as a perturbation to the $\hat{\sigma}_z$ eigenstates. According to second order perturbation theory, the perturbed ground state eigenenergy in the rotating frame~\footnote{Doing perturbation theory in a rotating frame does not give correct results in general. In this case it works, since the unperturbed Hamiltonian $\delta\hat{\sigma}_z/2$ commutes with the rotating-frame generator $\omega_L\hat{\sigma}_z/2$. For a compact state-centered derivation see e.g. Ref.~\cite{Foot:2005}, Ch.~7.7.} is
\begin{equation}
    E_0 \approx -\frac{\delta}{2} - \frac{\Omega^2}{4\delta} .
\end{equation}
For a fixed laser wavelength and thus detuning $\delta$, the decrease in energy is linear in the light intensity $I \propto |\vec{E}|^2 \propto \Omega^2$. Under the assumption that the electronic dynamics are much faster than the atom's motion, this intensity-dependent energy acts as a potential $U(x) = -\Omega(x)^2/(4\delta) \propto -I(x)$. Since the potential depends on the detuning $\delta$ and the dipole moment $\vec{\mu}$ of a specific transition, the potential which an atom experiences in general depends on its electronic state.

\section{Squeezing and Displacements in Optical Dipole Traps} \label{appendix: squeezing, displacement}

In this Appendix we derive the unitary operations that realize squeezing and conditional displacements of the harmonic oscillator states of an atom in an optical dipole trap.

\subsection{Squeezing} \label{app:Squeezing derivation}
The protocol shown in Fig.~\ref{fig:tweezer squeezing demonstration} realizes a unitary evolution that consists of a $\pi/2$ rotation in phase space followed by a squeezing in $\hat{p}$:
\begin{equation}
    \label{eq:Appendix squeezing unitary}
    \hat{V} = \hat{S} \Big(-\mathrm{ln}\left(\omega/\omega'\right) \Big) \; e^{-i \frac{\pi}{2} \hat{a}^\dag \hat{a}} .
\end{equation}
The operation takes the time $T = \frac{\pi}{2 \omega'}$.
Note that any idling oscillator will keep rotating at frequency $\omega$ during this time. This means that a phase difference of $\frac{\pi}{2} \left( \frac{\omega}{\omega'} - 1 \right)$ will build up between an oscillator following the described protocol and an idling oscillator.
For general durations $T$, a weaker squeezing is applied along a different axis, as can be seen in the following.
The Hamiltonian $\hat{H}=\hbar \omega \hat{a}^\dagger \hat{a}$ is changed diabatically to $\hat{H}' = \hbar \omega \hat{a}^\dagger \hat{a} + \frac{1}{2} m \left(\omega'^2-\omega^2\right)\hat{x}^2 = \hbar \omega \hat{a}^\dagger \hat{a} + \hbar \frac{\omega'^2-\omega^2}{2\omega}\hat{q}^2$. It can be expressed in terms of adjusted creation and annihilation operators: $H' = \hbar \omega' \hat{b}^\dagger \hat{b}$, where $\hat{b} = \cosh(r)\hat{a} - \sinh(r)\hat{a}^\dagger = \hat{S}^\dagger(r)\hat{a}\hat{S}(r)$ with the rapidity $r=\frac{1}{2}\ln\left(\omega/\omega'\right)$ and the temporary frequency $\omega'$. Letting the system evolve under this Hamiltonian for a time $T$ realizes the unitary operation
\begin{equation}
    \begin{aligned}
        \hat{V}(T) = & \; e^{-i\omega'T\hat{b}^\dagger\hat{b}} \\
        = & \; e^{-i\omega'T\hat{S}^\dagger(r)\hat{a}^\dagger\hat{a}\hat{S}(r)} \\
        = & \;\hat{S}^\dagger(r)e^{-i\omega'T\hat{a}^\dagger\hat{a}}\hat{S}(r) \\
        = & \; \hat{S}^\dagger(r)e^{-i\omega'T\hat{a}^\dagger\hat{a}}\hat{S}(r) e^{i\omega'T\hat{a}^\dagger\hat{a}} e^{-i\omega'T\hat{a}^\dagger\hat{a}} \\
        = & \; \hat{S}^\dagger(r) e^{\frac{r}{2}(e^{2i\omega'T}\hat{a}^2 - e^{-2i\omega'T}\hat{a}^{\dagger 2})} e^{-i\omega'T\hat{a}^\dagger\hat{a}} \\
        = & \; \hat{S}^\dagger(r)\hat{S}(re^{-2i\omega'T})e^{-i\omega'T\hat{a}^\dagger\hat{a}} \\
        = & \; \hat{S}(-r)\hat{S}(re^{-2i\omega'T})e^{-i\omega'T\hat{a}^\dagger\hat{a}} .
    \end{aligned}
\end{equation}
If the two subsequent squeeze operations act along the same axis in phase space, i.e.~if $T=\frac{\pi}{2 \omega'} =: T_0$, one achieves the maximal possible squeezing, arriving at Eq.~\eqref{eq:Appendix squeezing unitary}.

\paragraph*{Squeezing by fast modulation of the trap depth.}

\begin{figure*}
    \centering
    \includegraphics[width=0.99\linewidth]{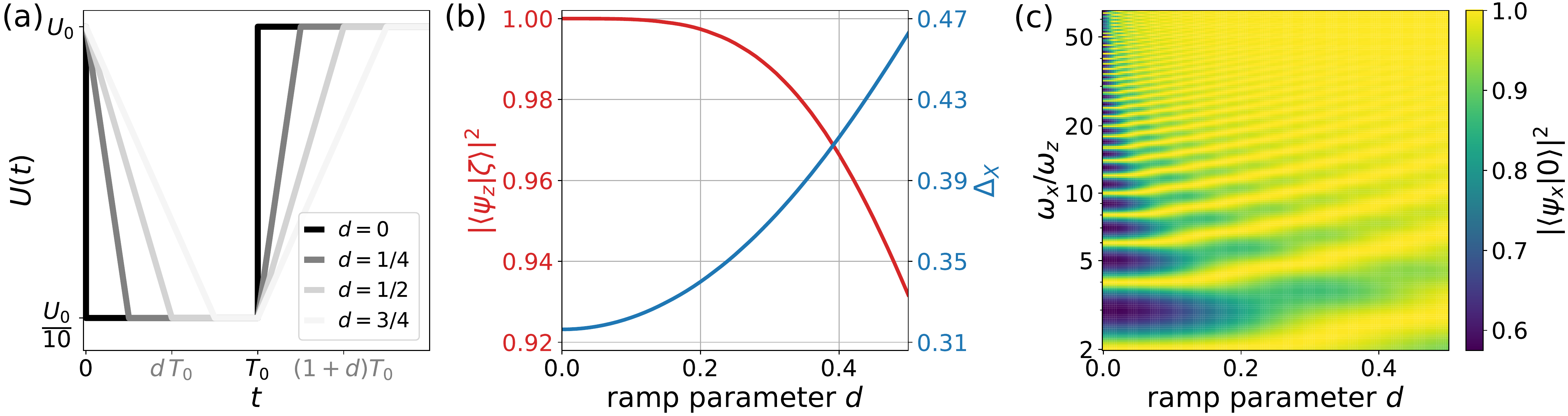}
    \caption{\textbf{Squeezing by fast modulation of the trap depth.} The motional state of an atom in an optical dipole trap that we model as a harmonic oscillator can be squeezed by a sudden change of the trap depth $U$ and thus a change of the oscillator frequency $\omega \propto \sqrt{U}$. (a) We consider linear ramps from the initial trap depth $U_0$ down to $U_0/10$. The time scale on which the potential changes is described by the ramp parameter $d$. (b) Effect of the squeezing procedure on the motional state $\ket{\psi_z}$ in the coding mode with frequency $\omega_z$. The instantaneous ramp protocol yields a state $\ket{\zeta} \coloneq \hat{S}(-\ln(10)/2)\ket{0}$ that has an effective squeezing $\Delta_X = 1/\sqrt{10} \approx 0.316$. The longer we choose the ramp time, the less squeezed is the resulting state. (c) Effect of the squeezing procedure on the motional state $\ket{\psi_x}$ in the spectator mode with frequency $\omega_x$. For larger ramp parameters $d$ as well as for larger oscillator frequency ratios $\omega_x/\omega_z$ the pulse appears more ``adiabatic'' and the final state remains closer to the ground state.}
    \label{fig:ramping}
\end{figure*}

As described in Sec.~\ref{sec:gkp_preparation} of the main text, the frequency $\omega$ of an atom in an optical dipole trap, which we describe as a harmonic oscillator, relates to the trap depth $U$ as $\omega \propto \sqrt{U}$. The motional state of the atom can thus be squeezed by a sudden change of the trap depth. An instantaneous change of the trap depth from $U_0$ to $U_0'$  yields a squeezing by $-\frac{1}{2}\ln(U_0/U_0')$ in a time $T_0 = \pi/(2 \omega')$, as described above.
The trapped atom exhibits three vibrational modes, however, we wish to squeeze just the motional state in encoded in the $z$-mode. The spectator modes $x$ and $y$ shall remain in their ground states.
In order to do so, one can ramp the trap depth at a speed that is fast compared to $\omega_z$ but slow in comparison with $\omega_x = \omega_y$. Here we consider, for simplicity, linear ramps from the initial trap depth $U_0$ down to $U_0' = U_0/10$. We describe the ramp speed with a parameter $d$ defined such that the ramp takes a time $d\,T_0$, as shown in Fig.~\ref{fig:ramping}(a). We remain in the adjusted oscillator frame for a time $(1-d)T_0$, which turns out to give good squeezing results. Eventually, the initial trap depth is restored by another linear ramp analogous to the one decreasing the trap depth.
Figure~\ref{fig:ramping}(b) shows the effect of the ramp duration on the motional state in the $z$-mode starting in the ground state. The instantaneous ramp yields the state $\ket{\zeta} \coloneq \hat{S}(-\ln(10)/2)\ket{0}$ that has an effective squeezing $\Delta_X = 1/\sqrt{10} \approx 0.316$. As expected, the resulting state $\ket{\psi_z}$ has smaller overlap with $\ket{\zeta}$ and larger effective squeezing $\Delta_X$ the larger we choose the ramp parameter $d$. Note that for $d \neq 0$ the squeezing axis of the final state is not parallel to the quadrature $q$-axis; however, including an appropriate idling time after the pulse will yield a momentum-squeezed state.
Figure~\ref{fig:ramping}(c) displays the effect of the ramp duration $d$ and the effect of the ratio of oscillator frequencies $\omega_x / \omega_z$ on the motional state in the spectator mode $x$. The color indicates the fidelity of the resulting state $\ket{\psi_x}$ w.r.t.~the initial ground state $\ket{0}$. For larger ramp parameters $d$ as well as for larger ratios of the oscillator frequencies $\omega_x / \omega_z$ the ramp time $d \, T_0$ gets larger as compared to the oscillator period $2\pi/\omega_x$. Therefore, the pulse appears more ``adiabatic'' and the resulting state remains closer to the ground state. Moreover, we see that for even values of $\omega_x/\omega_z$ the state can be kept in the ground state even at small values of $d$.

\subsection{Displacement} \label{app:Displacement derivation}
The Hamiltonian of a harmonic oscillator with an additional force term reads:
\begin{equation}
    H' = \frac{\hat{P}_x^2}{2m} + \frac{1}{2} m \omega^2 \hat{x}^2 - f \hat{x} .
\end{equation}
In terms of the quadrature operators this can be written as
\begin{equation}
    H' = \frac{1}{2} \hbar \omega \left( \hat{p}^2 + \left(\hat{q} - \alpha_d \right)^2 \right) - \frac{1}{2} \alpha_d^2 \hbar \omega,
\end{equation}
with $\alpha_d = f/\sqrt{\hbar m \omega^3}$.
We see that the harmonic oscillator is displaced by $\alpha_d$ in phase space, or by $d =\sqrt{\hbar/(m \omega)}\alpha_d = f/(m \omega^2)$ in real space. It also experiences an energy shift by $-\frac{1}{2} \alpha_d^2 \hbar \omega$ which manifests as a global phase when applied unconditionally and causes a phase-feedback if the displacement is conditioned on another qubit.
Thus, the Hamiltonian temporarily changes from $\hat{H}=\hbar\omega\hat{a}^\dagger\hat{a}$ to $\hat{H}'=\hbar\omega\hat{a}^\dagger\hat{a} - f\hat{x} = \hbar\omega \hat{b}^\dagger\hat{b} - \frac{1}{2} \alpha_d^2 \hbar \omega $ with $\hat{b} = \hat{a} - \alpha_d/\sqrt{2} = \hat{D}^\dagger(-\alpha_d)\hat{a}\hat{D}(-\alpha_d)$.
The evolution under this Hamiltonian for a time $t$ reduces to a combined rotation and displacement (see Fig.~\ref{fig:tweezer addition demonstration}):
\begin{equation}
\begin{aligned}
    \hat{U}(t) &= e^{-i\omega t\hat{b}^\dagger\hat{b} +\frac{i}{2} \alpha_d^2 \omega t} \\
    &= e^{-i\omega t\hat{D}^\dagger(-\alpha_d)\hat{a}^\dagger\hat{a} \hat{D}(-\alpha_d)+\frac{i}{2} \alpha_d^2 \omega t} \\
    &= \hat{D}^\dagger(-\alpha_d)e^{-i\omega t\hat{a}^\dagger\hat{a}+\frac{i}{2} \alpha_d^2 \omega t} \hat{D}(-\alpha_d) \\
    &= \hat{D}(\alpha_d) e^{-i\omega t\hat{a}^\dagger\hat{a}} \hat{D}(-\alpha_d) e^{i\omega t\hat{a}^\dagger\hat{a}} e^{-i\omega t\hat{a}^\dagger\hat{a}} e^{\frac{i}{2} \alpha_d^2 \omega t}  \\
    &= \hat{D}(\alpha_d)\hat{D}(-\alpha_d e^{-i\omega t}) e^{-i\omega t\hat{a}^\dagger\hat{a}} e^{\frac{i}{2} \alpha_d^2 \omega t}  \\
    &= \hat{D}\left( \alpha_d \left( 1 - e^{-i \omega t} \right) \right) e^{-i \omega t \hat{a}^\dag \hat{a}} e^{\frac{i}{2} \alpha_d^2 \omega t-\frac{i}{2}\alpha_d^2\sin(\omega t)} ,
\end{aligned}
\end{equation}
where we used that $e^{-i\omega t \hat{a}^\dagger \hat{a}} \hat{a} e^{i\omega t \hat{a}^\dagger \hat{a}} = e^{i \omega t} \hat{a}$ and $\hat{D}(\alpha)\hat{D}(\beta) = e^{-iA(\alpha,\beta)/2} \hat{D}(\alpha + \beta)$ with $A(\alpha, \beta) = \mathrm{Re}(\alpha) \mathrm{Im}(\beta) - \mathrm{Im}(\alpha) \mathrm{Re}(\beta)$.
One can achieve an effective displacement in $\hat{p}$ by letting the oscillator idle for a time $\frac{\pi}{\omega}-\frac{t}{2}$, then applying the unitary $\hat{U}(t)$ and letting the oscillator idle again for a time $\frac{\pi}{\omega}-\frac{t}{2}$. The whole procedure then takes a single oscillator period and realizes the evolution
\begin{equation}
\begin{aligned}
     &e^{-i(\pi - \omega t /2)\hat{a}^\dag \hat{a}} \hat{U}(t) e^{-i(\pi - \omega t /2)\hat{a}^\dag \hat{a}} \\
     = \; & e^{-i(\pi - \omega t /2)\hat{a}^\dag \hat{a}} \hat{D}(\alpha_d)e^{-i\omega t\hat{a}^\dagger\hat{a}+\frac{i}{2} \alpha_d^2 \omega t} \hat{D}(-\alpha_d) e^{-i(\pi - \omega t /2)\hat{a}^\dag \hat{a}} \\
     =\; &e^{-i\pi \hat{a}^\dag \hat{a}} \hat{D}(\alpha_de^{i\omega t/2})\hat{D}(-\alpha_de^{-i\omega t/2}) e^{-i\pi\hat{a}^\dag \hat{a}}e^{\frac{i}{2} \alpha_d^2 \omega t} \\
     =\; &e^{-i\pi \hat{a}^\dag \hat{a}} \hat{D}\left(2i\alpha_d\sin(\omega t/2)\right)e^{-\frac{i}{2}\alpha_d^2\sin(\omega t)} e^{-i\pi\hat{a}^\dag \hat{a}}e^{\frac{i}{2} \alpha_d^2 \omega t} \\
     =\; & \hat{D}\left(-2i\alpha_d \sin\left( \omega t / 2 \right) \right) e^{\frac{i}{2} \alpha_d^2 (\omega t-\sin(\omega t))}.
\end{aligned}
\end{equation}
Since the protocol takes a complete oscillator period, no phase space rotation compared to an idling oscillator builds up. Only the phase $\theta(t) = \frac{1}{2}\alpha_d^2(\omega t -\sin(\omega t))$ remains as an artifact on a conditioning qubit and can be corrected by a qubit-only rotation, since it does not depend on the bosonic state.

\section{Potentials of the Optical Tweezer and Optical Lattice} \label{appendix: Tweezer and Lattice Potentials}

In this Appendix we state the functional forms of the optical tweezer and optical lattice potentials that have been used in the main text. Moreover, we derive the oscillator frequencies, anharmonicities and couplings resulting from these potentials.

\subsection{Optical Tweezer}
An optical tweezer is formed from a tightly focused laser beam.
Under the paraxial approximation, the intensity of a TEM\textsubscript{00} mode of such a laser beam is described as follows~\cite{saleh_teich}:
\begin{equation}
    I(x,y,z) = I_0 \frac{1}{1+\frac{z^2}{z_R^2}}\exp(-2\frac{x^2+y^2}{w_0^2(1+\frac{z^2}{z_R^2})}),
\end{equation}
where $I_0$ is the peak intensity, $w_0$ is the beam width and $z_R = \pi w_0^2/\lambda$ is the Rayleigh length. A real laser, which does not have a pure TEM\textsubscript{00}-mode, has a higher beam divergence, which corresponds to a lower value of $z_R$. The tweezers considered in this paper reach the limits of the validity of the paraxial approximation, since $z_R \not\gg w_0$. Nonetheless, the Gaussian beam is considered a good estimate to extract the oscillator frequencies, anharmonicities and couplings.
Since the AC-Stark potential is proportional to the beam intensity, $U(x,y,z) = -U_0 \cdot I(x,y,z)/I_0$, the potential energy up to fourth order in the position coordinates reads as follows:
\begin{equation} \label{eq:tweezer_taylor}
     U(x,y,z) = U_0 \bigg(-1 + \frac{z^2}{z_R^2} + \frac{2(x^2+y^2)}{w_0^2}
    - \frac{z^4}{z_R^4} - \frac{2(x^4+y^4)}{w_0^4}
        - \frac{4z^2(x^2+y^2)}{z_R^2 w_0^2} - \frac{4x^2y^2}{w_0^4} \bigg) + \mathcal{O}\left((x^2,y^2,z^2)^3\right).
\end{equation}
Identifying the coefficients in front of the harmonic terms $x^2$, $y^2$ and $z^2$ with $\frac{1}{2} m \omega_{x,y,z}^2$ yields the following relations for the oscillator frequencies:
\begin{equation}
        \omega_z^2 = \frac{2 U_0}{m z_R^2}, \qquad
        \omega_{x}^2 = \omega_{y}^2 = \frac{4 U_0}{m w_0^2} .
\end{equation}
Using these expressions, we can write Eq.~\eqref{eq:tweezer_taylor} as follows:
\begin{equation}
    U = -U_0 + \frac{m}{2} \left( \omega_z^2 z^2 + \omega_x^2 (x^2 + y^2) \right) - \frac{m^2}{U_0} \left( \frac{\omega_z^4}{4} z^4 + \frac{\omega_x^4}{8} (x^4 + y^4) + \frac{\omega_z^2 \omega_x^2}{2} z^2 (x^2 + y^2) + \frac{\omega_x^4}{4} x^2 y^2 \right) + \mathcal{O}\left((x^2,y^2,z^2)^3\right) .
\end{equation}
Changing from real space positions to quadratures $q_j = j \sqrt{m \omega_j/\hbar}$, for $j \in \{ x, y, z\}$ we arrive at
\begin{equation}
    U = -U_0 + \frac{\hbar}{2} \left( \omega_z q_z^2 + \omega_x (q_x^2 + q_y^2) \right) - \frac{\hbar^2}{U_0} \left( \frac{\omega_z^2}{4} q_z^4 + \frac{\omega_x^2}{8} (q_x^4 + q_y^4) + \frac{\omega_z \omega_x}{2} q_z^2 (q_x^2 + q_y^2) + \frac{\omega_x^2}{4} q_x^2 q_y^2 \right) + \mathcal{O}\left((q_x^2,q_y^2,q_z^2)^3\right) .
\end{equation}
Comparing this equation with
\begin{equation}
    \frac{U}{\hbar} = -\frac{U_0}{\hbar} + \frac{1}{2} \left( \omega_z q_z^2 + \omega_x (q_x^2 + q_y^2) \right) - \eta_z \omega_z q_z^4 - \eta_x \omega_x (q_x^4 + q_y^4) - \varepsilon_{zx} \omega_z q_z^2 (q_x^2 + q_y^2) - \varepsilon_{xy} \omega_x q_x^2 q_y^2 + \mathcal{O}\left((q_x^2,q_y^2,q_z^2)^3\right)
\end{equation}
we can identify the anharmonicities and couplings
\begin{equation}
        \eta_z = \frac{\hbar \omega_z}{4 U_0}, \quad
        \eta_x = \frac{\hbar \omega_x}{8 U_0}, \quad
        \varepsilon_{zx} = \frac{\hbar \omega_x}{2 U_0}, \quad
        \varepsilon_{xy} = \frac{\hbar \omega_x}{4 U_0} .
\end{equation}

\subsection{2D Optical Lattice}
The 2D-cross-section of a rectangular folded optical lattice can be described as follows, assuming infinitely wide beams~\cite{Wei:2023, Tao:2024}:
\begin{equation}
\begin{aligned}
    I(x,y) = \frac{I_0}{8}
    \Biggl(& 2+ \cos(\frac{4\pi}{\lambda} (x \cos\Theta + y \sin\Theta))
    + \cos(\frac{4\pi}{\lambda} (x \cos\Theta - y \sin\Theta)) \\
    & + 2 \cos(\frac{4\pi}{\lambda} y \sin\Theta)
    + 2 \cos(\frac{4\pi}{\lambda} x \cos\Theta)\Biggr) .
\end{aligned}
\end{equation}
The angle $\Theta$ can be adjusted to change the lattice geometry, e.g. for $\Theta=45^\circ$ one obtains a square lattice. The intensity drop-off from finitely wide beams can be modelled by a Gaussian envelope in the $z$-direction. The beam width in the $x$- and $y$-directions can be made much larger than the lattice sites, such that this intensity drop-off can be neglected for an individual central lattice site but has to be reconsidered when using very large lattices.
For $\Theta=45^\circ$ the potential reads
\begin{equation} \label{eq:lattice_taylor}
\begin{aligned}
    U(x,y,z) = & -\frac{U_0}{8} \exp(-2 z^2/w_0^2) \Biggl( 2+\cos(\frac{4\pi}{\sqrt{2}\lambda} (x + y )) + \cos(\frac{4\pi}{\sqrt{2}\lambda} (x  - y )) + 2 \cos(\frac{4\pi}{\sqrt{2}\lambda} y ) + 2 \cos(\frac{4\pi}{\sqrt{2}\lambda} x)\Biggr)\\[0.2cm]
    = & U_0 \left( -1 + \frac{2z^2}{w_0^2} + \frac{2\pi^2(x^2+y^2)}{\lambda^2} - \frac{2z^4}{w_0^4}  - \frac{4\pi^4(x^4+y^4)}{3\lambda^4} - \frac{4\pi^2z^2(x^2+y^2)}{\lambda^2w_0^2} - \frac{4\pi^4x^2y^2}{\lambda^4}\right) + \mathcal{O}\left((x^2,y^2,z^2)^3\right) , 
\end{aligned}
\end{equation}
where $U_0$ is the trap depth.
The oscillator parameters can be obtained analogously to the tweezer analysis in the previous subsection.
Identifying the coefficients in front of the harmonic terms $x^2$, $y^2$ and $z^2$ with $\frac{1}{2} m \omega_{x,y,z}^2$ yields the following relations for the oscillator frequencies:
\begin{equation}
        \omega_z^2 = \frac{4 U_0}{m w_0^2}, \qquad
        \omega_{x}^2 = \omega_{y}^2 = \frac{4 \pi^2 U_0}{m \lambda^2} .
\end{equation}
Using these expressions, we can write Eq.~\eqref{eq:lattice_taylor} as follows:
\begin{equation}
    U = -U_0 + \frac{m}{2} \left( \omega_z^2 z^2 + \omega_x^2 (x^2 + y^2) \right) - \frac{m^2}{U_0} \left( \frac{\omega_z^4}{8} z^4 + \frac{\omega_x^4}{12} (x^4 + y^4) + \frac{\omega_z^2 \omega_x^2}{4} z^2 (x^2 + y^2) + \frac{\omega_x^4}{4} x^2 y^2 \right) + \mathcal{O}\left((x^2,y^2,z^2)^3\right) .
\end{equation}
Changing from real space positions to quadratures $q_j = j \sqrt{m \omega_j/\hbar}$ for $j \in \{ x, y, z\}$ we arrive at
\begin{equation}
    U = -U_0 + \frac{\hbar}{2} \left( \omega_z q_z^2 + \omega_x (q_x^2 + q_y^2) \right) - \frac{\hbar^2}{U_0} \left( \frac{\omega_z^2}{8} q_z^4 + \frac{\omega_x^2}{12} (q_x^4 + q_y^4) + \frac{\omega_z \omega_x}{4} q_z^2 (q_x^2 + q_y^2) + \frac{\omega_x^2}{4} q_x^2 q_y^2 \right) + \mathcal{O}\left((q_x^2,q_y^2,q_z^2)^3 \right) .
\end{equation}
Comparing this equation with
\begin{equation}
    \frac{U}{\hbar} = -\frac{U_0}{\hbar} + \frac{1}{2} \left( \omega_z q_z^2 + \omega_x (q_x^2 + q_y^2) \right) - \eta_z \omega_z q_z^4 - \eta_x \omega_x (q_x^4 + q_y^4) - \varepsilon_{zx} \omega_z q_z^2 (q_x^2 + q_y^2) - \varepsilon_{xy} \omega_x q_x^2 q_y^2 + \mathcal{O}\left((q_x^2,q_y^2,q_z^2)^3\right)
\end{equation}
we can identify the anharmonicities and couplings
\begin{equation}
        \eta_z = \frac{\hbar \omega_z}{8 U_0}, \quad
        \eta_x = \frac{\hbar \omega_x}{12 U_0}, \quad
        \varepsilon_{zx} = \frac{\hbar \omega_x}{4 U_0}, \quad
        \varepsilon_{xy} = \frac{\hbar \omega_x}{4 U_0} .
\end{equation}

\bibliography{refs}

\end{document}